\documentclass[a4paper]{article}
\usepackage{authblk}
\usepackage[utf8]{inputenc}
\usepackage{amssymb,amsmath,amsthm}
\usepackage{url}
\usepackage{multirow}
\usepackage{enumitem}
\usepackage{graphicx}
\usepackage{algorithm}
\usepackage[noend]{algpseudocode}
\usepackage{float}
\usepackage{booktabs}
\usepackage{numprint}
\usepackage{todonotes}
\usepackage{tabularx}
\usepackage{ragged2e}
\usepackage[group-separator={,}]{siunitx}
\setlist{nolistsep}

\newcommand{\startlist}{
\setlength{\itemsep}{0mm}
\setlength{\topsep}{ 0mm}\setlength{\leftmargin}{4mm}
\setlength{\rightmargin}{0mm}
}
\newcommand{\listlabel}{$\bullet$}

\usepackage[graphicx]{realboxes}
\usepackage{subfigure}

\makeatother

\date{\vspace{-5ex}}

\newif\ifarxiv
\arxivtrue

\begin{document}
\title{A Cryptoeconomic Traffic Analysis of Bitcoin's Lightning Network}

\ifarxiv
\author[1,2]{Ferenc B\'eres}
\author[2]{István A. Seres}
\author[1,3]{Andr\'as A. Bencz\'ur}
\affil[1]{Institute for Computer Science and Control (SZTAKI), Hungary}
\affil[2]{E\"otv\"os University, Budapest, Hungary}
\affil[3]{Széchenyi University, Gy\H{o}r, Hungary}
\else
\author{}
\fi

\maketitle

\begin{abstract}
Lightning Network (LN) is designed to amend the scalability and privacy issues of Bitcoin. It is a payment channel network where Bitcoin transactions are issued off the blockchain and onion routed through a private payment path with the aim to settle transactions in a faster, cheaper, and more private manner, as they are not recorded in a costly-to-maintain, slow, and public ledger. In this work, we design a traffic simulator to empirically study LN's transaction fees and privacy provisions.
The simulator relies only on publicly available data of the network structure and capacities, and generates transactions under assumptions that we attempt to validate based on information spread by certain blog posts of LN node owners.

Our findings on the estimated revenue from transaction fees are in line with the widespread opinion that participation is economically irrational for the majority of the large routing nodes who currently hold the network together. Either traffic or transaction fees must increase by orders of magnitude to make payment routing economically viable.  We give worst-case estimates for the potential fee increase by assuming strong price competition among the routers.  We also estimate how current channel structures and pricing policies respond to a potential increase in traffic, how reduction in locked funds on channels would affect the network, and show examples of nodes who are estimated to operate with economically feasible revenue.

Our second set of findings considers privacy.  Even if transactions are onion routed, strong statistical evidence on payment source and destination can be inferred, as many transaction paths only consist of a single intermediary by the side effect of LN's small-world nature.  Based on our simulation experiments, we (1) quantitatively characterize the privacy shortcomings of current LN operation, and (2) propose a method to inject additional hops in routing paths to demonstrate how privacy can be strengthened with very little additional transactional cost.
\end{abstract}


\section{Introduction} \label{sec:introduction}
Bitcoin is a peer-to-peer, decentralized cryptographic currency~\cite{nakamoto2008bitcoin}. It is a censorship-resistant, permissionless, digital payment system. Anyone can join and leave the network whenever they would like to. Participants can issue payments, which are inserted into a distributed, replicated ledger called blockchain. Since there is no trusted central party to issue money and guard this financial system,  payment validity is checked by all network participants. The necessity of full validation severely limits the scalability of decentralized cryptocurrencies: Bitcoin could theoretically process $27$ transactions per second (tps)~\cite{georgiadis2019many}; however, in practice its average transaction throughput is $7$ tps~\cite{croman2016ds}. This is in stark contrast with the throughput of mainstream payment providers; for example, in peak hours Visa is able to achieve $\num[group-separator={,}]{47000}$ tps on its network~\cite{trillo2013stress}.

To alleviate scalability issues, the cryptocurrency community is continuously inventing new protocols and technologies. A major line of research is focused on amending existing currencies without modifying the consensus layer by introducing a new layer, i.e.,\ off-chain transactions~\cite{mccorry2016towards,miller2017sprites, dziembowski2017perun}. These proposals are called Layer-2 protocols: they allow parties to exchange transactions locally, without broadcasting them to the blockchain network, updating a local balance sheet instead and only utilizing the blockchain as a recourse for disputes. For an exhaustive review of off-chain protocols, refer to~\cite{gudgeon2019sok}. 

Among these proposals, the most prominent ones are payment channel networks (PCN), in which nodes have several open payment channels, being able to connect to all nodes, possibly through multiple hops. The most popular instantiation of a PCN is Bitcoin's Lightning Network (LN)~\cite{poon2016bitcoin}, a public, permissionless PCN, which allows anyone to issue Bitcoin transactions without the need to wait for several blocks for payment confirmation and currently with transaction fees orders of magnitude lower than on-chain fees. LN is suitable for several application scenarios, for instance, micropayments or e-commerce, with the intent to make everyday Bitcoin usage more convenient and frictionless. LN's core value proposition is that Bitcoin users can send low-value payments instantly in a privacy-preserving manner with negligible fees, which has led to quite a widespread adoption of LN among Bitcoin users.

The main difficulty with analyzing how LN operates is that  the exact transaction routes are cryptographically hidden from eavesdroppers due to onion routing~\cite{kate2010using}.
LN can only be observed through public information on nodes and channel openings, closings, and capacity changes. The actual amount of Bitcoins circulated in LN is unknown, although in blog posts, some node owners publish high-level statistics, such as their revenue~\cite{lnbig_post,bitmex_post}, which can be used as grounds for estimation. 

To analyze LN efficiency and profitability, we designed a \textbf{traffic simulator} for LN to analyze the routing costs and potential revenue at different nodes.  We assigned roles to nodes by collecting external data\footnote{Source: \url{https://1ml.com}}, labeling nodes as wallet services, shops, and other merchants. Using node labels, we simulated the flow of Bitcoin transactions from ordinary users towards merchants over time, based on the natural assumption that transactions are routed through the path that charges the minimum total transaction fee. By taking the dynamically changing transaction fees of the LN nodes into account, we designed a method to predict the optimal fee pricing policy for individual nodes in case of the cheapest path routing.

To the best of our knowledge, there has been no previous \textbf{empirical study on LN  transaction fees}. Our traffic simulator hence opens the possibility for addressing questions of transaction routes, amounts, fees, and other measures otherwise depending upon strictly private information, based solely on the observable network structure. 
By releasing the source code of our tool, we allow node owners to fit various parameters to their private observation on LN traffic.
In particular, in this paper the simulator enables us to draw two major conclusions:
\begin{description}
\item[Economic incentives.] Currently, LN provides little to no financial incentive for payment routing. Low routing fees do not sufficiently compensate the routing nodes that essentially hold the network together. Our results show that in general, transaction fees are underpriced, since for many possible payments there is no alternative path to execute the transaction. We also give estimates of how the current network and fee structure responds to increase in traffic and decrease in channel capacities, thus assessing the income potential in different strategies. We provide an open source tool for nodes to experimentally design their channels, capacities, and fees by incorporating all possible information that they privately infer from the traffic over their channels.
\item[Privacy.] We quantitatively analyze the privacy provisions of LN. Despite onion routing, we observe that strong statistical evidence can be gathered about the sender and receiver of LN payments, since a substantial portion of payments involve only a single routing intermediary, who can easily de-anonymize participants. 
We find that using deliberately suboptimal, longer routing paths can potentially restore privacy while only marginally increasing the cost of an average transaction, as it is partially already incorporated in other implementations of the Lightning protocol~\cite{grunspan2018ant}.
\end{description}{}

The rest of the paper is organized as follows. In Section~\ref{sec:relatedworks}, we review the growing body of literature on PCNs and specifically on LN. In Section~\ref{sec:background}, we provide a brief background on LN and its fee structure. In Section~\ref{sec:trafficSimulation}, our traffic simulator is presented. We discuss our experimental results in three sections. We investigate the price competition and the potential to increase fees, under various assumptions, in Section~\ref{sec:competition}.
We estimate the profitability of the central router nodes under estimated current and potentially increased future traffic in Section~\ref{sect:profitablity}.
Finally, we estimate the amount of privacy shortcomings due to too short paths and potential mitigations in Section~\ref{sec:privacy}. We conclude our paper in Section~\ref{sec:conclusion}.

\section{Related Works} \label{sec:relatedworks}
To the best of our knowledge, we have conducted the first empirical analysis on LN transaction fees, similar to the way empirical and theoretical studies on on-chain transaction fees have been conducted during the early adoption of cryptocurrencies. Möser and Böhme conducted a longitudinal study on Bitcoin's nascent transaction fee market~\cite{moser2015trends}. Kaskaloglu asserted that near-zero transaction fees cannot last long as block rewards diminish~\cite{kaskaloglu2014near}. Easley et al.\ developed a game-theoretic model to explain the factors leading to the emergence of transactions fees, and provided empirical evidence on the model predictions~\cite{easley2019mining}. 
Recently, BitMEX, a single LN node, has experimented with setting different transaction fees to measure the effect on routing revenue~\cite{bitmex_post}, which shows a similar pattern to our simulation experiments.

Unlike on-chain transactions, the LN transaction fee market is not yet consolidated. Some actors behave financially rationally, while the vast majority exhibit altruistic behavior, which parallels the early days of Bitcoin~\cite{moser2015trends}. Similarly to on-chain fees, we expect to see more maturity and a similar evolution in the LN transaction fee market in the future.

Even before the launch of LN, many works studied the theoretical aspects of PCNs. Branzei et al.\ studied the impact of LN on Bitcoin transaction costs~\cite{branzei2017charge}. They conjectured a lower miner income from on-chain transaction fees as users tend to use and issue transactions on LN.
In~\cite{khan2019lightning}, the transaction fees of various payment channels are compared, however, without reference to the underlying network dynamics.

Depleted payment channels account for many efficiency issues in PCNs. Khalil and Gervais devised a handy algorithm to revive imbalanced payment channels without opening new ones~\cite{khalil2017revive}. 

PCNs can also be considered to be creation games. A user might decide to create a payment channel to a destination node or just route the payment in the already existing PCN. The former is more expensive; however, repeated payments can amortize the on-chain cost of opening a payment channel. Avarikioti et al.\ found that given a free routing fee policy, the star graph constitutes a Nash equilibrium~\cite{avarikioti2019payment}. In a similar game-theoretic work, the effect of routing fees was analyzed~\cite{avarikioti2018payment}. It was again found that the star graph is a near-optimal solution to the network design problem. 

Even though transactions in LN are not recorded on the blockchain, they do not provide privacy guarantees. As early as 2016, Herrera et al.\ anticipated the privacy issues emerging in a PCN~\cite{herrera2016privacy}. Single-intermediary payments do not provide privacy, although they have higher utility. Tang et al.\ asserts that a PCN either operates in a low-privacy or a low-utility regime~\cite{tang2019privacyutility}. Although a recently devised cryptographic protocol solves the privacy issues of single-intermediary routed payments~\cite{tairi2}, the protocol is not yet in use due to its complexity of implementation.

After the launch of LN, several studies have investigated the graph properties of LN~\cite{seres2019topological,rohrer2019discharged,martinazzi2019evolution}. They described the topology of LN at an arbitrarily chosen point in time and found that LN exhibits a hub and spoke topology, and its degree distribution can be well approximated with a scale-free distribution~\cite{seres2019topological,rohrer2019discharged}. Furthermore, these works assessed the robustness of the network against various types of attack strategies: they showed that LN is susceptible to both node~\cite{seres2019topological,martinazzi2019evolution} and channel~\cite{rohrer2019discharged} removal based attacks. These works are restricted to a static snapshot of LN. The lack of temporal data has largely limited the insights and results of these contributions.

In a Youtube video~\cite{earn2019pickhardt}, an estimate of the routing income is given based on the assumption that the payment probability between any node pair is the same. As it is easy to see, under this assumption the routing income of a node is proportional to its betweenness centrality. In our simulation experiments, we will explicitly compare our prediction with the one based on betweenness centrality and show how the finer structure of our estimation procedure yields more plausible results.

At the time of writing, four research groups published results on payment channel network simulators, each serving purposes very different from ours.
Out of them, the simulator of Branzei et al.~\cite{branzei2017charge} is the only one that has pointers to publicly available resources.  Their simulator only considers single bidirectional channels or a star topology, and its main goal is to analyze channel opening costs and depletion.
This simulator is extended in~\cite{engelmann2017towards} to generate and analyze Barab\'asi-Albert graphs as underlying networks.
CLoTH~\cite{conoscenti2018cloth} is able to provide performance statistics (e.g.,\ probability of payment failure on a given PCN graph); however, it does not analyze transaction fees, profitability, optimal fee policy, and privacy provisions of LN. In contrast, our LN traffic simulator can produce insights in those areas as well. 
Finally, the simulator in~\cite{zhang2019cheapay} is a distributed method to minimize the transaction fee of a payment path, subject to the timeliness and feasibility constraints for the success ratio and the average accepted value of the transactions.

\section{Routing and Fees in Lightning Network Payment Channels} \label{sec:background}

A \textbf{payment channel} allows users to make multiple cryptocurrency transactions without committing all of the transactions to the  blockchain. In a typical payment channel, only two transactions are added to the blockchain, but theoretically, an unlimited number of payments can be made between the participants. Parties can open a payment channel by escrowing funds on the blockchain for subsequent use only between those two parties.  The sum of the individual balances on the two sides of the channel is usually referred to as the \textbf{capacity}.

We illustrate the operation of a payment channel by an example. Let Alice and Bob  escrow 1 and 2 tokens respectively, by committing a transaction to the blockchain that sets up a new channel. 
Once the channel is finalized, Alice and Bob can send escrowed funds back and forth by revoking the previous state of the channel and digitally signing the new state updated by the  transacted tokens. For example, Alice can send 0.1 of her 1 token to Bob, so that the new channel state is (Alice=0.9, Bob=2.1). Once the parties decide to close the channel, they can commit its final state through another blockchain transaction. 

Maintaining a payment channel has an opportunity cost since users must lock up their funds while the channel is open, and funds are not redeemable until the channel is closed. Hence, it is not practical to expect users to maintain a channel with every individual with whom they may ever need to transact. 

In a \textbf{payment channel network} (PCN), nodes have several open payment channels between each other; however, not necessarily with all other nodes. The network of bidirectional payment channels allows two parties to exchange funds even if they do not have a direct payment channel. For example, if Alice has a balance of 1 token with Ingrid, and Ingrid has a balance of 2 tokens with Bob locked in a payment channel, then Alice can route payments to Bob through Ingrid up to the maximum of the balances of Alice and Ingrid. Assuming that Alice sends 0.2 tokens to Bob, after routing we have the following channel balances: Alice=0.8, Ingrid=0.2 on the first channel and Ingrid=1.8, Bob=0.2 on the second channel. 

In a payment channel, cryptographic protections are used to ensure that channel updates in both directions are executed atomically, i.e.,\ either both or neither of them are performed~\cite{gudgeon2019sok}. In addition, incentive-based protections are also implemented to prevent users from stealing funds in a channel, e.g., by committing a revoked state. 
Similar techniques allow payment routing for longer paths. Furthermore, payment router intermediaries are financially motivated to relay payments as they are entitled to claim transaction fees after each successfully routed payment.
 
LN as a PCN consists of nodes representing users and undirected, weighted edges representing payment channels. Users can open and close bidirectional payment channels between each other and route payments through these connections.  Therefore, LN can be modeled as an undirected, weighted multigraph since nodes can have multiple channels between each other. The weights on the edges correspond to the capacity of the payment channels. 

In LN only capacities of payment channels are known publicly, individual balances are kept secret. This is because if individual balances are known, balance updates would reveal successful transactions, hence preventing transaction privacy.

\begin{figure}[ht!]
\centering
\begin{minipage}{.45\textwidth}
    \centering
    \includegraphics[width=\linewidth,trim={0cm 0cm 0cm 0cm},clip]{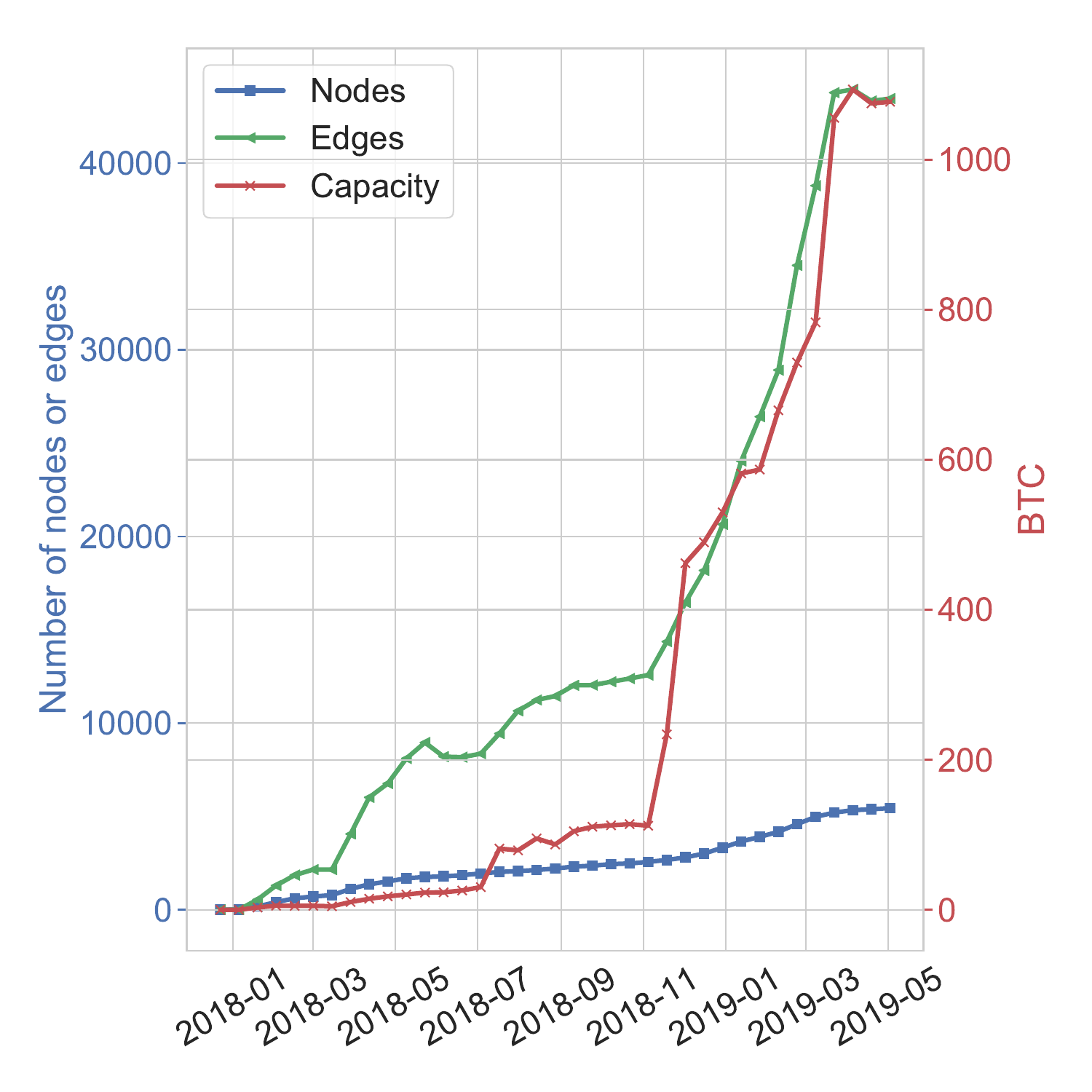}
    \caption{LN's increasing popularity and adoption in its first 17 months.}
    \label{fig:lnovertime}
\end{minipage}
\hspace{6pt}
\begin{minipage}{.45\textwidth}
    \centering
    \includegraphics[width=\linewidth]{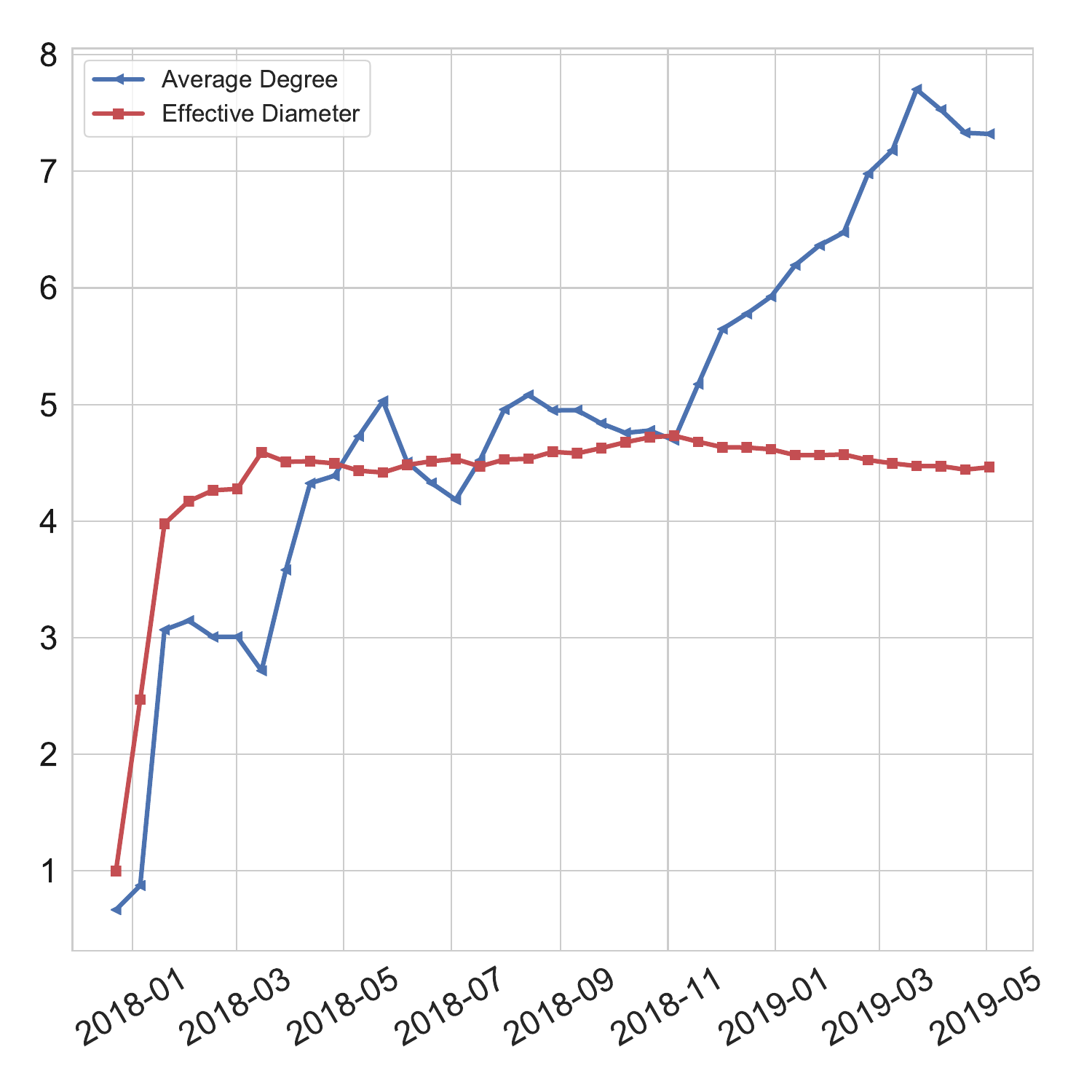}
    \caption{Average degree and effective diameter in LN, as the function of time. }
    \label{fig:avgDegreeEffectiveDiameter}
\end{minipage}
\end{figure}

\subsection{Routing in LN and Fee Mechanism}\label{sec:routing_fee}
LN applies source routing, meaning that it is always the sender who decides the payment route towards the intended recipient. Packets are onion routed, which means that intermediary nodes only know the identity of their immediate predecessor and successor in the route.  Therefore, from a privacy perspective, nodes are incentivized to avoid single-intermediary paths, as in those cases intermediaries are potentially able to identify both the sender and the receiver.

LN provides financial incentives for intermediaries to route payments. In LN there are two types of fees that a sender pays to the intermediaries in case the transaction involves more than one payment channels. Nodes can set and charge the following fees after each routed payments:
\begin{description}
\item[Base fee:] a fixed fee denoted as \text{baseFee}, charged each time a payment is routed through the channel.
\item[Fee rate:] a percentage fee denoted as \text{feeRate}, charged on the value \text{txValue} of the payment.  
\end{description}
Therefore, the total transaction fee \text{txFee} to an intermediary can be obtained as: 
\begin{equation}
\text{txFee} = \text{baseFee} + \text{feeRate}\cdot\text{txValue}.
\label{eq:tx_fee}
\end{equation}
We note that the base fee and fee rate is set by individual users, thus forming a fee market for payment routing. Furthermore, we remark that Equation~\ref{eq:tx_fee} does not hold for all routing algorithms. However, we do not consider other fee structures in our simulator, as currently alternative routing algorithms are not widely adopted throughout the network.

\subsection{Data}
\label{sec:data}

Throughout our work, we analyze two main data sources that are both available online\ifarxiv\footnote{See:~\url{https://github.com/ferencberes/LNTrafficSimulator}}\fi.
First, we gathered an edge stream data that describes every payment channel opening and closure from block height \num[group-separator={,}]{501337} (in December 28, 2017) to \num[group-separator={,}]{576140} (in May 15, 2019).  Second, we collected snapshots of \emph{the public graph} using the \textit{lnd} client and utilized snapshots taken by Rohrer et al~\cite{rohrer2019discharged} as well. We highlight that only the latter dataset contains transaction fee information. Thus, the experiments in Sections~\ref{sec:trafficSimulation}-\ref{sec:privacy} are only based on $40$ consecutive LN graph snapshots from 2019 February and March.

We note that according to some estimates, $28\%$ of all channels are private~\cite{private2020bitmex}, meaning that their existence can only be recognized by the two ends. In our analysis, we have no information about private payment channels; however, the same holds for all the other network participants as well. Hence, we do not expect a significant bias in our results, as presumably those channels have private use and do not participate in carrying the global network traffic.

We labeled LN nodes by relying on the tags provided by the node owners\footnote{Source:~\url{https://1ml.com}}. This allows us to distinguish between ordinary users and \textbf{merchants}. We assume that merchants receive payments more often than regular users. This is essential in understanding how popular payment channels are depleted throughout LN by repeated use in one direction.
The number of merchant nodes in the union of all 40 snapshots is 169.

First we describe the graphs defined based on the $40$ consecutive LN graph snapshots from 2019 February and March.
We consider a minimum meaningful capacity $\alpha=60\,000$ (approximately USD 5) and exclude  edges with capacity less than $\alpha$ in $G$ as they cannot be used in payments with value $\alpha$.\footnote{Note that at the time of writing, atomic multipath payments (AMPs) are not implemented. AMPs would allow one to split a payment value into multiple smaller amounts and subsequently send those payments to the receiver via multiple payment paths through different intermediaries. The AMP protocol will guarantee that either all sub-payments are executed or none of them.}  Although LN channels are bidirectional, in our experiments we consider two directed edges, so that we can use channels in one direction if the capacity is exhausted in the other direction.  We also ignore edges in the direction where they are flagged as disabled in the data.  The properties of the LN network, averaged over the 40 daily snapshots, is as follows:
\begin{itemize}
    \item Number of the union of all nodes: 4\,787;
    \item Average number of nodes in a day: 3\,358;
    \item Non-isolated nodes after filtering disabled edge directions and edges with capacity less than 60\,000\,SAT: 3\,132;
    \item Size of the largest strongly connected component: 2\,206;
\end{itemize}

\begin{figure}
\centering
\begin{minipage}{.45\textwidth}
    \centering
    \includegraphics[width=\linewidth,trim={0cm 0cm 0cm 0cm},clip]{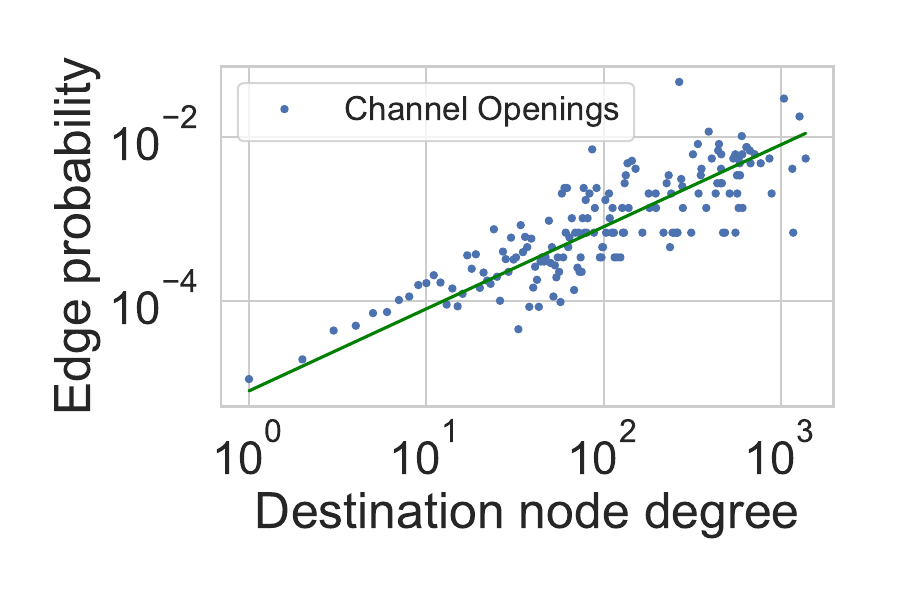}
    \caption{Preferential attachment in LN. The higher a node's degree, the higher the probability that it receives a payment channel.}
    \label{fig:preferential}
\end{minipage}
\hspace{6pt}
\begin{minipage}{.45\textwidth}
    \centering
    \includegraphics[width=0.95\textwidth]{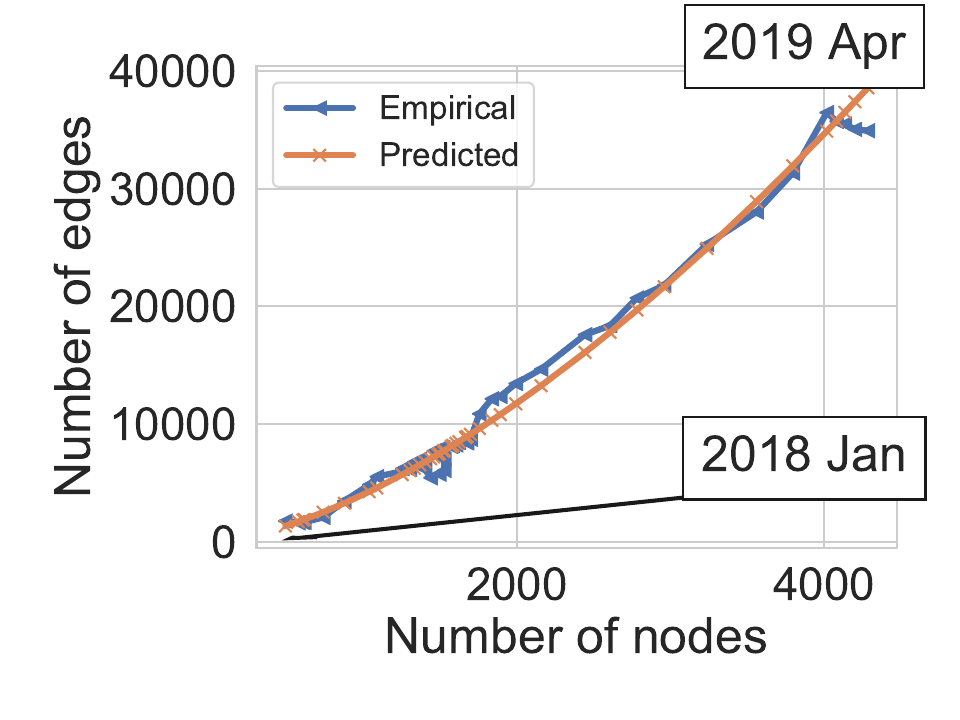}
	\caption{LN follows the Densification Power Law relation with exponent $a=1.55634117$. Goodness-of-fit: $R^2=0.98$.  }
	\label{fig:densification}
\end{minipage}
\end{figure}

The degree distribution of LN follows power law. The effect of preferential attachment, the phenomenon that new edges tend to attach to high degree nodes, is clearly seen in Figure~\ref{fig:preferential}. 
Ever since LN was launched, its popularity has grown steadily (Figure~\ref{fig:lnovertime}). This growth in popularity has caused the average degree increasing and the diameter decreasing over time, a ``densification'' phenomenon observed for a wide class of general networks in~\cite{leskovec2005graphs}. 
The average degree steadily increases, while the effective diameter decreases only after a first initial expansion phase (Figure~\ref{fig:avgDegreeEffectiveDiameter}), following the densification power law (Figure~\ref{fig:densification}).

\begin{figure}
\centering
\begin{minipage}{.45\textwidth}
    \centering
    \includegraphics[width=0.95\linewidth]{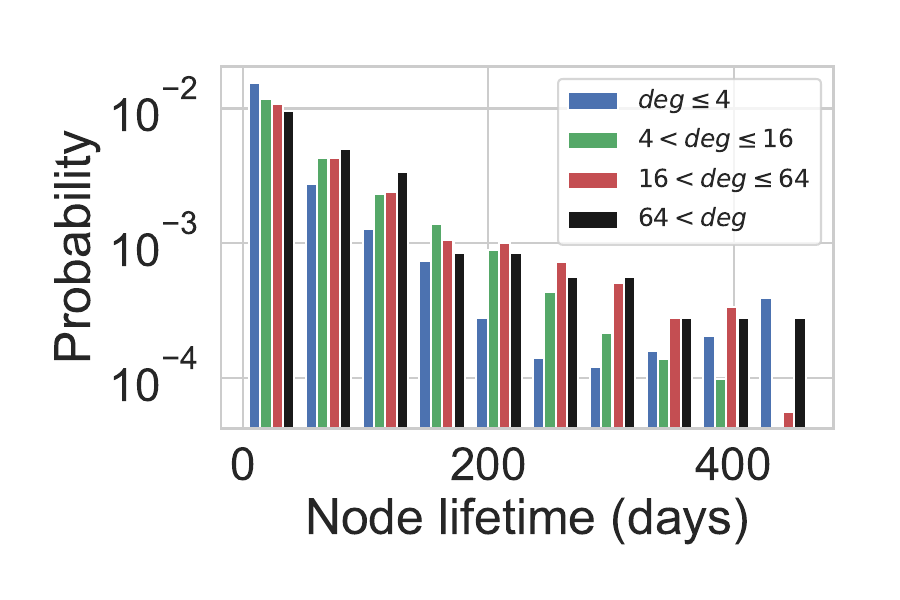}
    \caption{Node lifetime distribution in days, separately for four node degree groups.}
    \label{fig:nodeLifeTime}
\end{minipage}
\hspace{6pt}
\begin{minipage}{.45\textwidth}
    \centering
    \includegraphics[width=\linewidth]{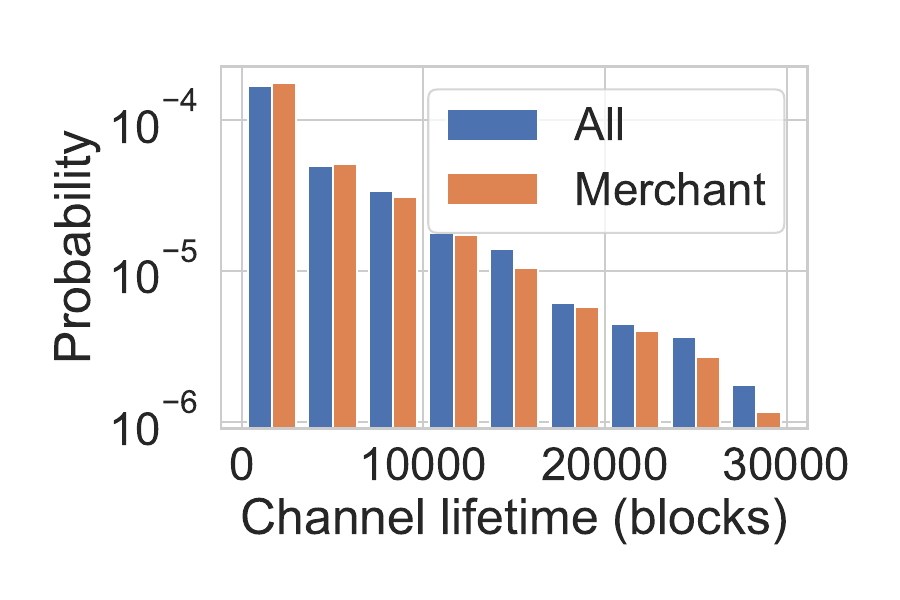}
    \caption{Channel lifetime distribution of merchants and others (merchant average: 5198; overall average: 5474).}
    \label{fig:merchantChannelLifeTime}
\end{minipage}
\end{figure}

\begin{figure}
\centering
\begin{minipage}{.45\textwidth}
    \centering
    \includegraphics[width=\linewidth]{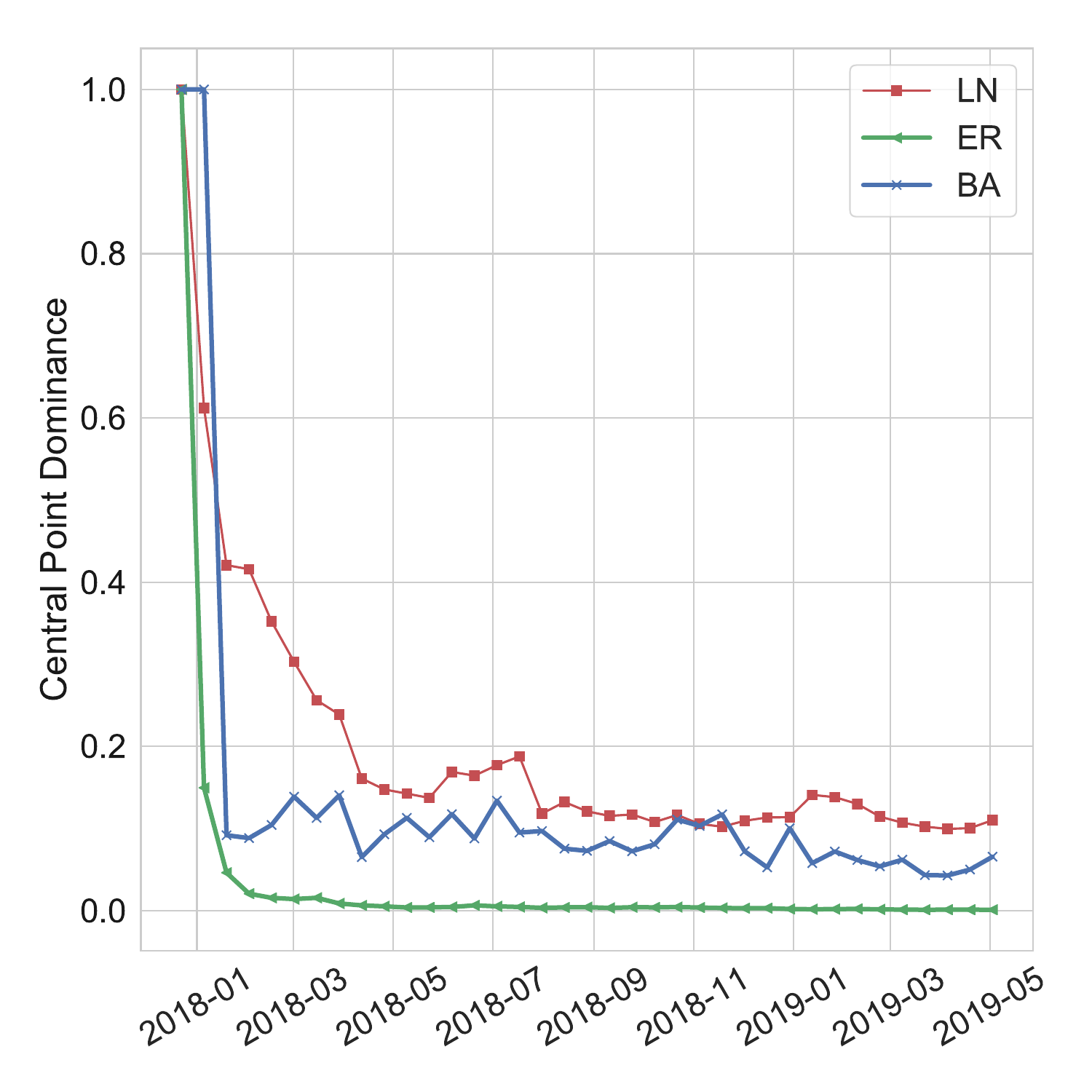}
    \caption{Central Point Dominance of LN as the function of time, compared to that of an Erdős-Rényi (ER) and a Barabási-Albert (BA) graph of equal size at the given time.}
    \label{fig:lnCPD}
\end{minipage}
\hspace{6pt}
\begin{minipage}{.45\textwidth}
    \centering
    \includegraphics[width=0.95\linewidth]{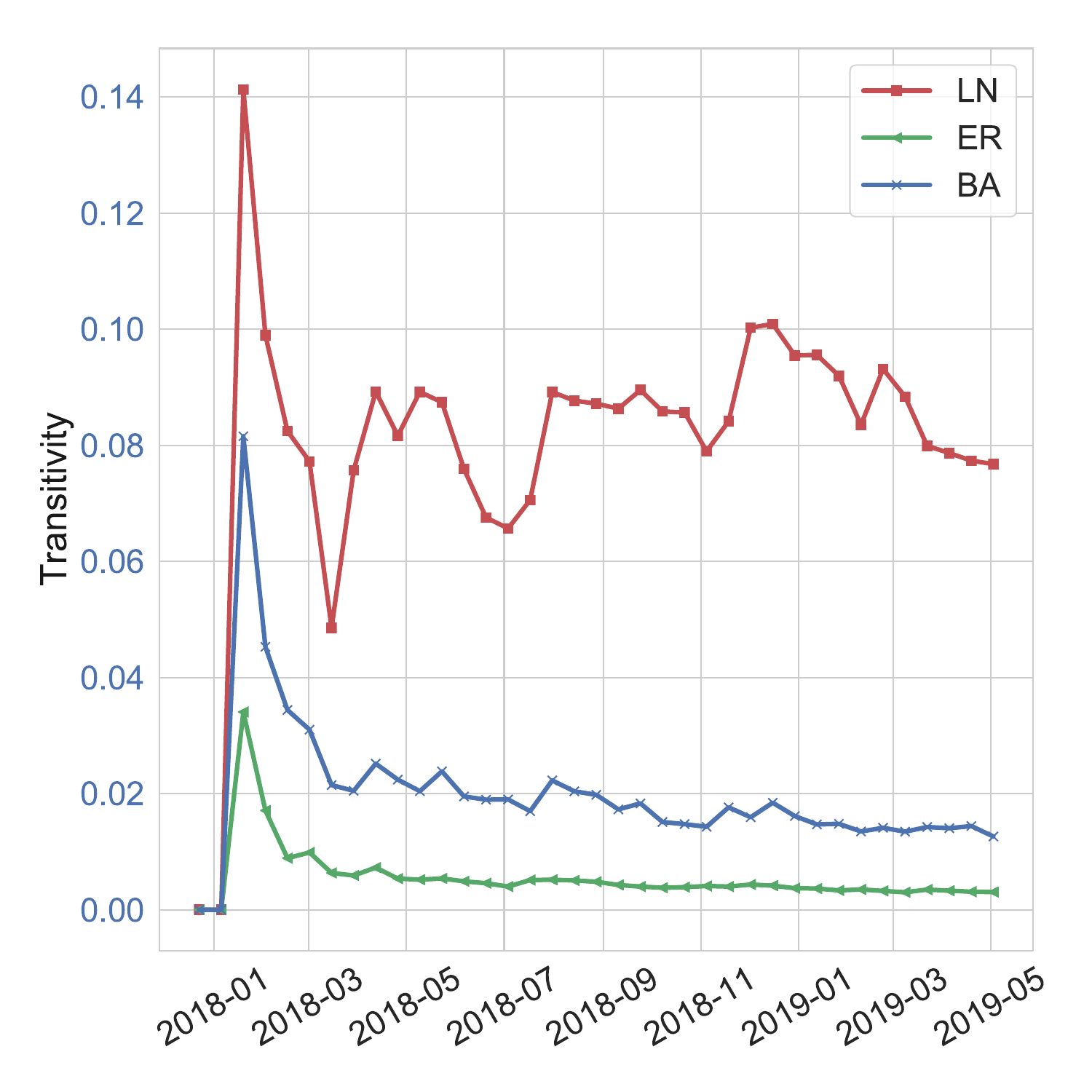}
    \caption{Transitivity of LN, compared to that of an Erdős-Rényi (ER) and a Barabási-Albert (BA) graph of equal size at the given time.}
    \label{fig:lnTransitivity}
\end{minipage}
\end{figure}

\begin{figure}
\centering
\includegraphics[width=0.4\linewidth,trim={0.3cm 0.9cm 0cm 0cm},clip]{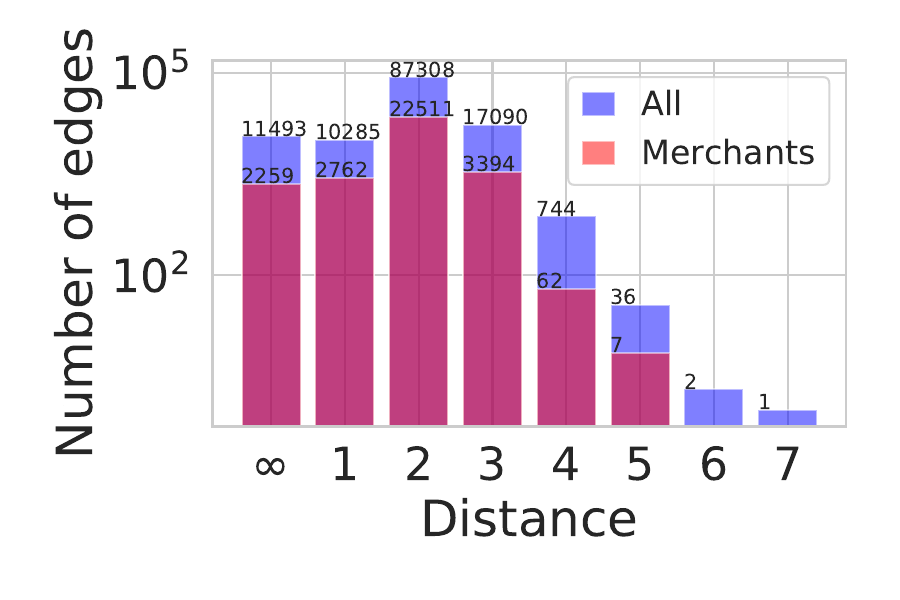}
\caption{The distance of LN nodes in the network at the time before a payment channel is established between them, shown separate for all nodes and for merchants only. If nodes were in different connected components before establishing a payment channel between them, then we define their distance as $\infty$.}
\label{fig:edgelocality}
\end{figure}

We observe that the higher its degree, the longer a node participate in LN, see Figure~\ref{fig:nodeLifeTime}.
Additionally, the channels adjacent to merchants have a shorter average lifetime (5198 blocks) than the average channel lifetime (5474 blocks), see the difference of the full distribution in Figure~\ref{fig:merchantChannelLifeTime}. We suspect that subsequent payments deplete the channels of the merchants, who then close these channels, collect their funds, and open new channels. 

We observe strong central point dominance in LN (Figure~\ref{fig:lnCPD}), which indicates that LN is more centralized than a Barabási-Albert or an Erdős-Rényi graph of equal size. This is in line with the predictions of~\cite{avarikioti2019payment,avarikioti2018payment}, affirming that PCNs lean to form a star graph like topology to achieve Nash equilibrium. 

Counterintuitively, LN also exhibits high transitivity, also known as global clustering coefficient, see Figure~\ref{fig:lnTransitivity}. One would expect that nodes have no incentive to close triangles, as they might as well just route payments along already existing payment channels. However, we observe that the vast majority ($68.76\%$) of all created payment channels connect nodes only 1 hop (distance 2) away from each other, see Figure~\ref{fig:edgelocality}. We believe that in most cases this is caused by replacing depleted payment channels. 
The high transitivity in LN is especially striking when it is compared to other social graphs. LN has roughly the same clustering coefficient as the YouTube social network~\cite{mislove2007measurement}.

\section{Lightning Network Traffic Simulator} \label{sec:trafficSimulation}

In this section, we introduce our main contribution, the LN Traffic Simulator, which we designed for daily routing income and traffic estimation of network entities.  Simulation is necessary to analyze the fine-grained structure, since the key concept of LN is privacy:  data will never include transaction amounts, sources, and targets in any form, and it is very unlikely that it will give information on the capacity distribution over the channels, since that would leak information on the actual transactions.  Hence we need a simulator to understand the capabilities and limitations of the network to route transactions.

By simulating transactions at different traffic volumes and  transaction amounts, we shed light on the fee pricing policies of major router entities as well as on privacy considerations, as we will describe in Sections~\ref{sec:competition}--\ref{sec:privacy}.

In our simulator, we make the assumption that the sender nodes always choose the cheapest route to execute their transactions.
Due to the source routing nature of LN, nodes are expected to possess the knowledge of network structure and current transaction fees to make price optimal decisions. 
Note that in the LN client\footnote{See \url{https://github.com/lightningnetwork/lnd} and \url{https://github.com/ElementsProject/lightning}.}, the source node selects the routing for their transactions. For example, the sender node may choose the shortest instead of the cheapest path to the target if speed is more important than the transaction cost, and our simulator can be modified accordingly.

The main goal of our traffic simulator is to generate a certain number of transactions, given as an input parameter, by using only the information on the edges and their capacities in a given LN snapshot. 
To generate transaction sources and targets, we predefine the fraction of the transactions that lead to merchants based on the assumption that the majority of the transactions correspond to money spent at shops and service providers. We fix the amount as constant to reduce the complexity of the simulation model. 

We acknowledge that using constant payment amounts is a strong assumption. One could consider various distributions such as Pareto, power law, Poisson, as in previous works~\cite{tang2019privacyutility}. However, assumptions on the distributions as well as their parameter settings greatly increase the complexity of the experimentation, and cannot be empirically validated, since payment values are not public.
We found the necessity to incorporate correlations of the amounts with node sizes and roles particularly troublesome.
We note that constant amounts are also capable of capturing larger values by repeated payments from the same node. Finally, any time some entities obtain reliable estimates on the payment value distribution, they can conduct the corresponding experiments with our open source simulator.

Formally, we use the following notation:
\begin{itemize}
    \item $G$, a daily graph snapshot of the LN with channels represented by pairs of edges in both directions; disabled directions and too low capacity edges are excluded;
    \item $M$, the set of merchant nodes defined in Section~\ref{sec:data};
    \item $\tau$, the number of random transactions to sample;
    \item $\alpha$, the (constant) value of each transaction, in satoshis\footnote{Each Bitcoin (BTC) is divisible to the 8th decimal place, so each BTC can be split into 100,000,000 units. Each unit of Bitcoin, or 0.00000001 Bitcoin, is called a Satoshi. A satoshi is the smallest unit of Bitcoin, see \url{https://satoshitobitcoin.co/}.};
    \item $\epsilon$, the ratio of merchants in the endpoints of the random transactions.
\end{itemize}

The available data only includes the total channel capacity but not its distribution between the endpoints. Thus, before simulation we randomly initialize the capacity between the channel endpoints. For example, if $\Gamma$ is the total capacity of the channel between nodes $u$ and $v$, we let $0\leq \gamma(uv)\leq \Gamma$ and $0\leq \gamma(vu)\leq \Gamma$ denote the maximum value in satoshis, which can be routed from $u$ to $v$ and vice versa. Both $\gamma(uv)$ and $\gamma(vu)$ change after each transaction that uses this channel while maintaining $\gamma(uv)+\gamma(vu)=\Gamma$ at all times.

If an edge has capacity less than $\alpha$  in a direction, that is $\gamma(uv)<\alpha$, the edge direction $uv$ is \emph{depleted}.  In the simulation, a depleted edge $uv$ cannot be used before a payment is made in the opposite direction $vu$, in which case $\gamma(uv)\geq\alpha$ will hold.  Optionally, in Section~\ref{sect:profitablity}, we will also investigate the effect of removing this constraint and allow the simulation to use an edge direction without limits.  We also note that routers can balance payment channels without closing and reopening existing ones by finding cycles containing a depleted channel and route funds on a circular payment path~\cite{khalil2017revive}, however, this option is not implemented in the current version of our simulator. 

We start the simulation by first sampling $\tau$ transactions, each of amount $\alpha$. First we select $\tau$ senders uniformly at random from all nodes.  Recipients are selected by putting emphasis on merchants $M$: we choose $\epsilon\cdot\tau$ merchants with probability proportional to their degree in addition to $(1-\epsilon)\cdot\tau$ recipients that are selected uniformly at random from all nodes including both merchants and non-merchants.
Finally, we randomly match senders and recipients.

Given the transactions, we are ready to simulate traffic by finding the cheapest paths $P=(s=u_0, u_1, u_2, \dots, u_k=t)$ from sender $s$ to recipient $t$ with the capacity constraint $\gamma(u_{i}u_{i+1}) \geq \alpha$ for $i=0\dots k-1$. Then, node statistics (e.g.,\ routing income, number of routed transactions) are updated for each intermediary node $\{u_1, u_2, \dots, u_{k-1}\}$ with respect to the latest transaction. Finally, for $i=0\dots k-1$ the value of $\gamma(u_{i}u_{i+1})$ is decreased while $\gamma(u_{i+1}u_{i})$ is increased by the transaction amount $\alpha$ in order to keep available node capacities up to date. As we work with daily graph snapshots, the simulation mimics the daily traffic on LN.

The simulated routing income of a node will arise as the sum of the payment costs of its inbound channels.
The cost of a payment can be obtained by substituting txValue${}=\alpha$ in the transaction fee Equation~(\ref{eq:tx_fee}), we obtain the transaction fee of an edge as $\text{baseFee} + \text{feeRate}\cdot\alpha$.  We note that in this work we give no estimate on the cost of opening the channels, instead, we stop using depleted edges as long as a payment in the opposite direction reactivates them.  We will assess the effect of channel depletion on routing income in Section~\ref{sect:profitablity}, where we will allow the simulation to use an edge direction without capacity limits.

Due to several random factors in the simulation, including source and target sampling and capacity distribution initialization, we run the traffic simulator ten times.  We use 40 consecutive daily snapshots in our data. We always report the mean node statistics (e.g.,\ node routing income, daily traffic) of LN entities over our sets of $400$ simulations for each parameter setting.

\subsection{Feasibility Validation and Choice of Parameters}
\label{sec:param_tuning}

We validate our simulation model by comparing published information with our estimates for the income and traffic of the most relevant LN router entities. These nodes are responsible for keeping the network operational by routing most of the transactions. Our key source of information is the blog post~\cite{lnbig_post} on LNBIG.com, the most relevant routing entity who owns several nodes on LN as well as approximately half of the total network capacity: 
\begin{itemize}
    \item In a typical day, LNBIG.com serves $200$--$300$  transactions through all of its nodes, rarely exceeding $600$ in a single day.
    \item On routing commissions, LNBIG.com earns $5,000$--$10,000$ satoshis per day.
\end{itemize}

\begin{figure}[ht!]
    \centering
    \includegraphics[width=0.75\linewidth]{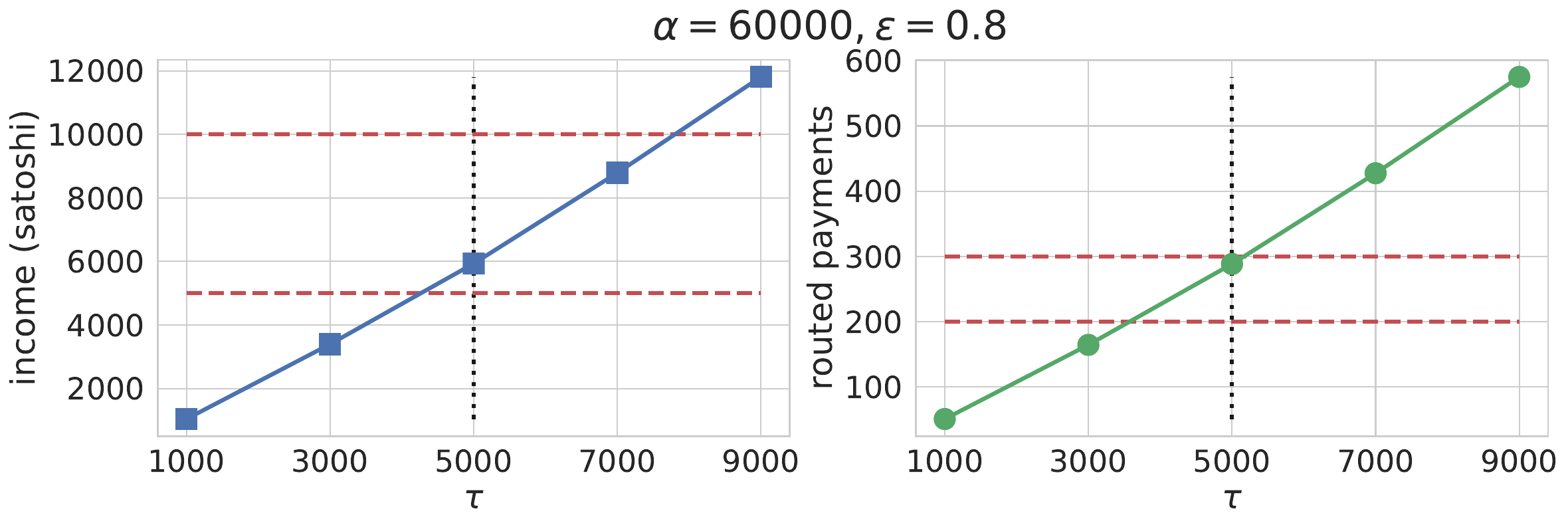}
    \includegraphics[width=0.75\linewidth]{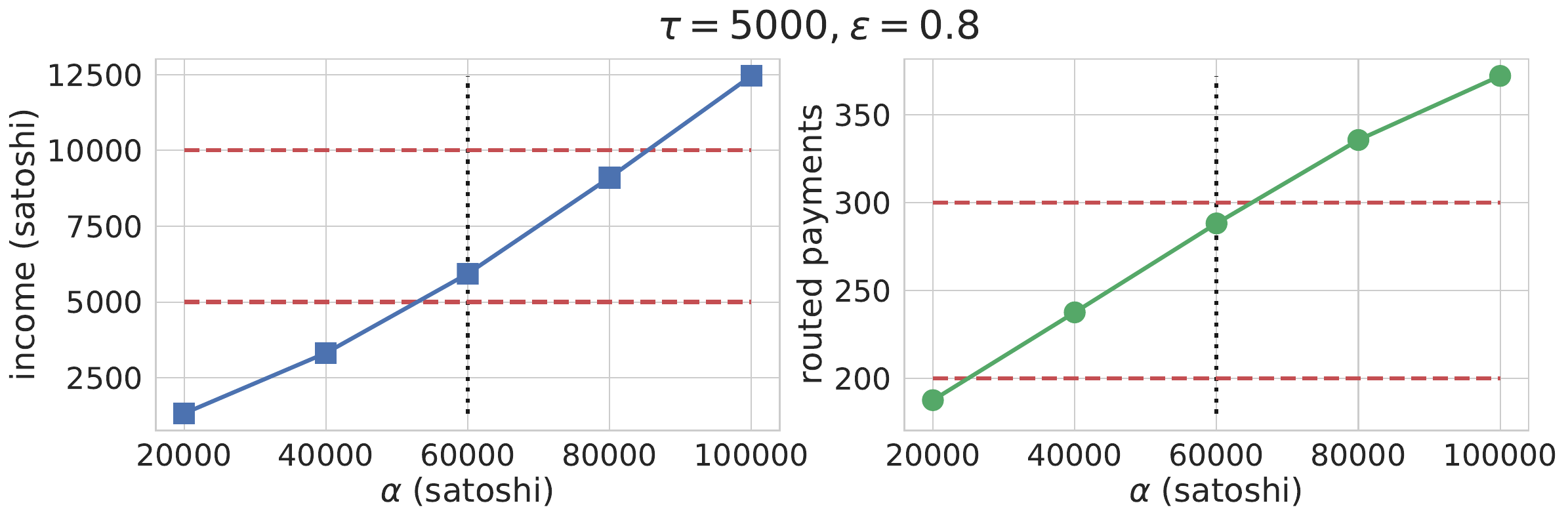}
    \includegraphics[width=0.75\linewidth]{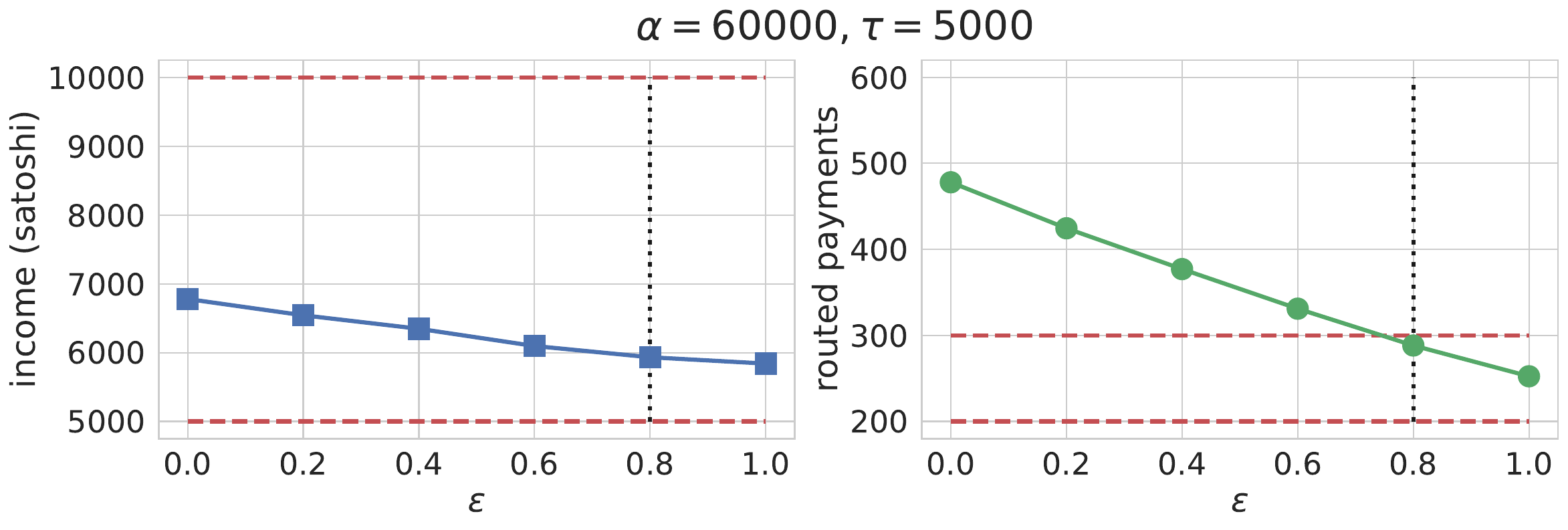}
    \caption{Mean estimated routing income and number of routed payments of LNBIG.com entity with respect to traffic simulator parameters.
    The default parameter setting (daily transaction count $\tau=5000$, single transaction amount $\alpha=60,000$ satoshis, and merchant endpoint ratio $\epsilon=0.8$) is marked by vertical black dotted lines.
    The daily income and traffic ranges stated by LNBIG.com~\cite{lnbig_post} are marked by horizontal red dashed lines. }
    \label{fig:parameter_tuning}
\end{figure}

We managed to reproduce daily traffic and routing income similar to LNBIG.com by sampling $\tau=5,000$ transactions with $\alpha=60,000$ satoshis (approximately 5 U.S.\ dollars) and merchant ratio $\epsilon=0.8$. The estimated revenue, as the function of the parameters, is shown in Figure~\ref{fig:parameter_tuning}, also showing the target daily income and traffic ranges stated by LNBIG.com~\cite{lnbig_post}.

To summarize, simulating a few thousand micro-payments with mostly merchant recipients resulted in similar traffic and revenue as described over the nodes of LNBIG.com.
We choose $\tau=5,000$, $\alpha=60,000$, and $\epsilon=0.8$ as default parameters of our traffic simulator in order to draw some conclusions on LN node profitability and transaction privacy in  Sections~\ref{sec:competition}--\ref{sec:privacy}.

\subsection{Traffic Simulator Response to Parameter Changes}

Next we examine the stability of our traffic simulator for different ratios of merchant endpoints  $\epsilon$. We note that the set of transaction recipients can be sampled uniformly at random by choosing $\epsilon=0.0$, while in case $\epsilon=1.0$, every sampled transaction has merchant endpoints. Thus, by increasing the value of $\epsilon$ the traffic can be centralized towards LN service providers. 
As determined in the previous subsection, we set the remaining parameters $\tau=5,000$ and $\alpha=60,000$.

\begin{figure}[ht!]
    \centering
    \subfigure[Correlation of routing traffic for $\epsilon \in \{0.0,0.2,0.5,0.8,1.0\}$.]{
    \includegraphics[width=0.3\linewidth]{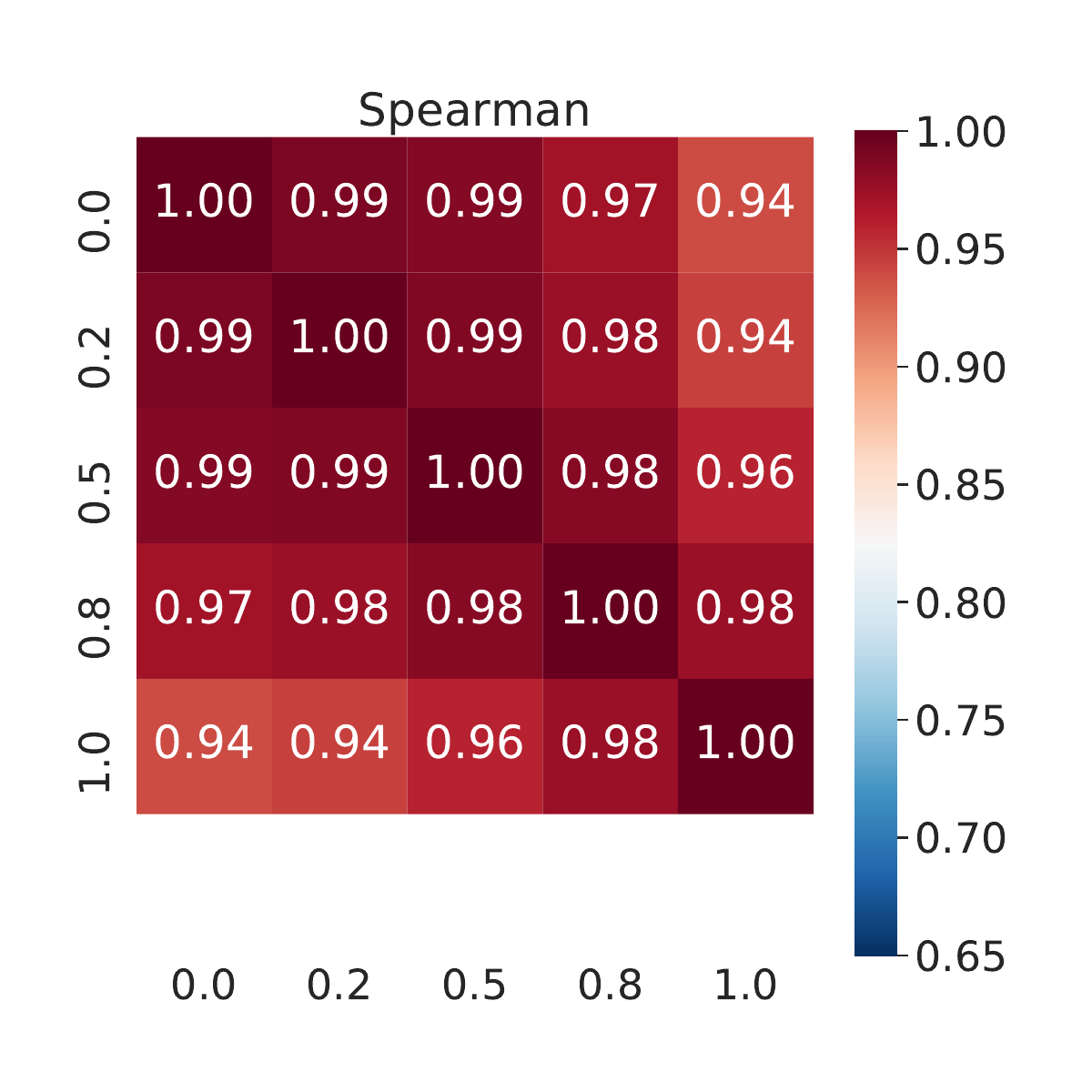}
    \includegraphics[width=0.3\linewidth]{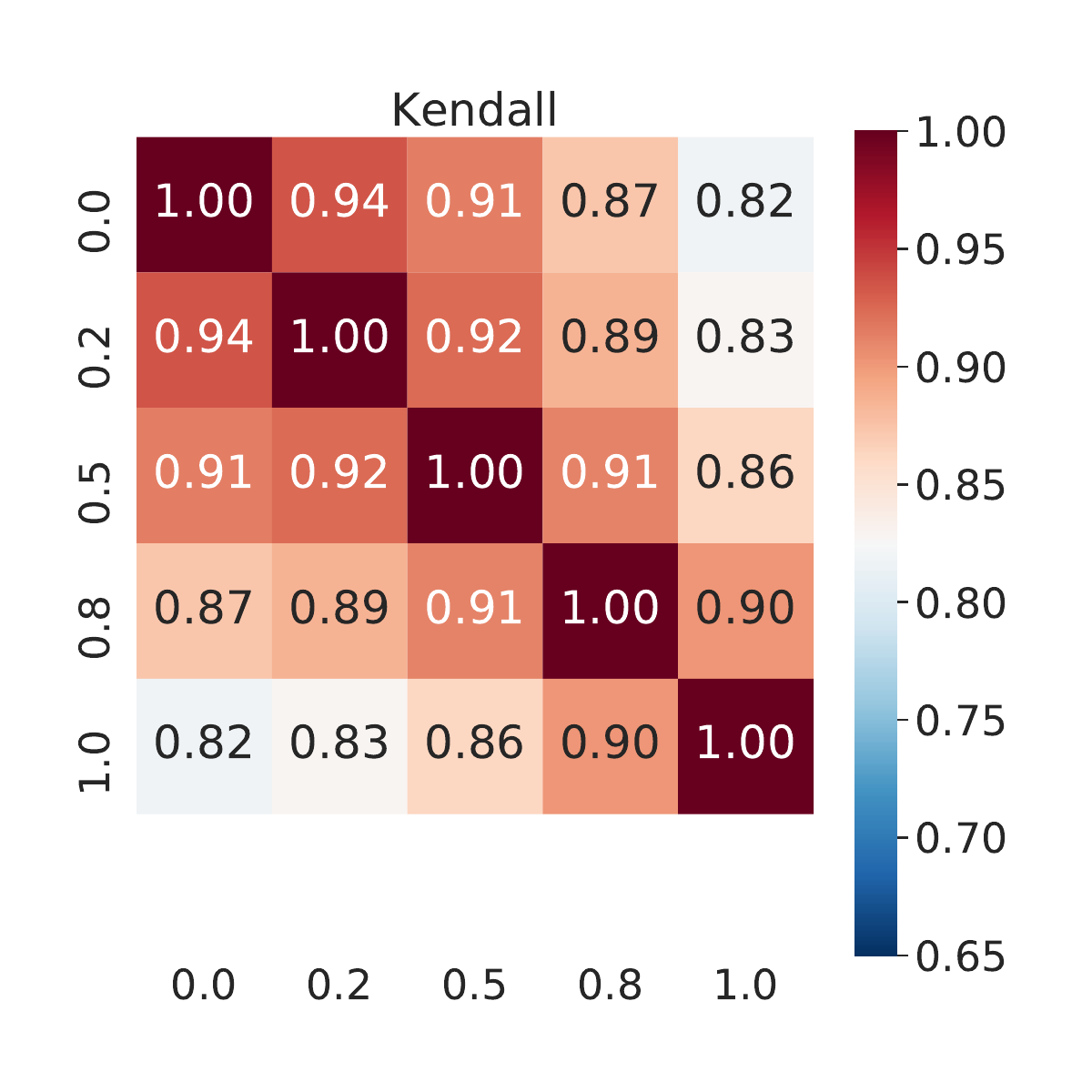}
    \includegraphics[width=0.3\linewidth]{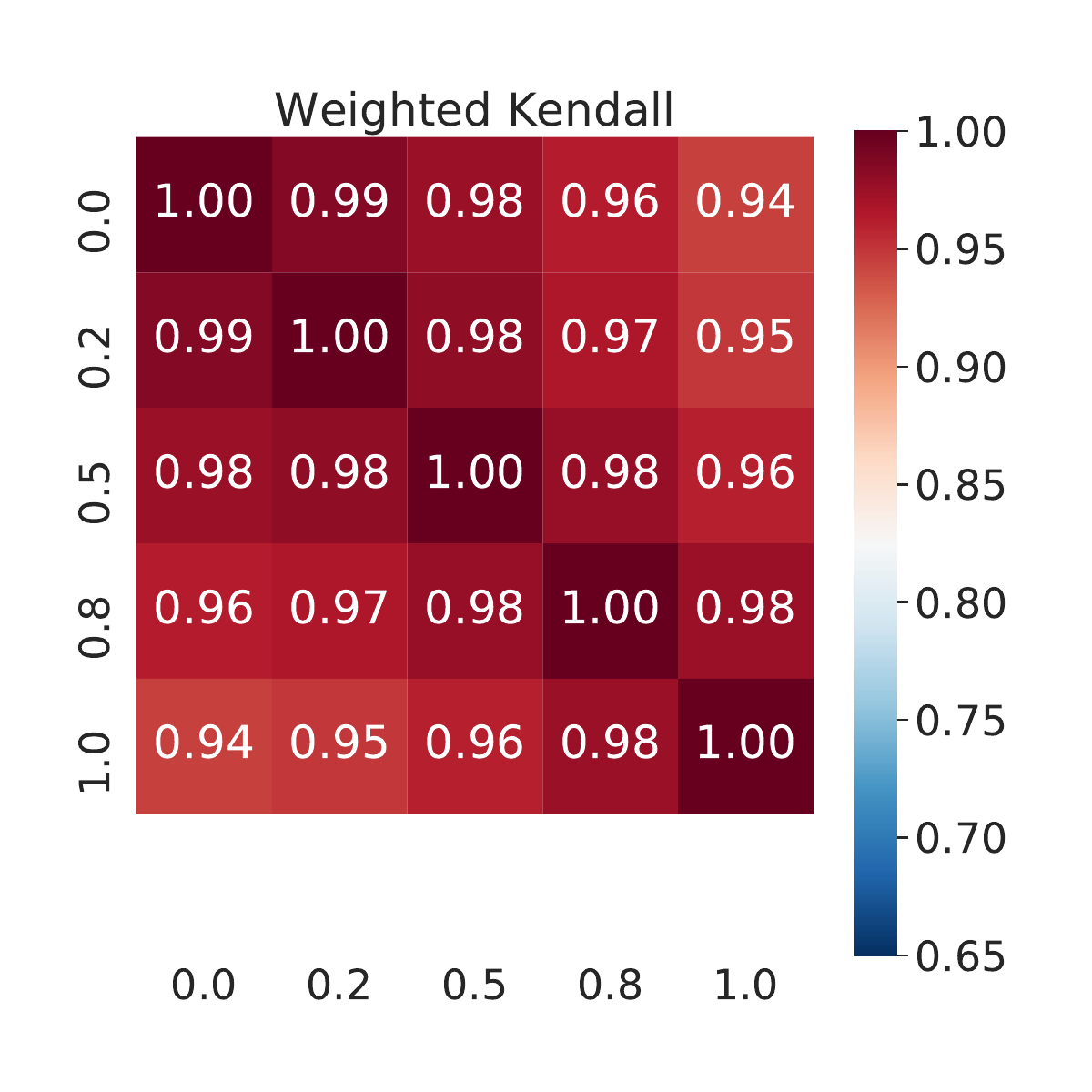}}
    \centering
    \subfigure[Correlation of routing income for $\epsilon \in \{0.0,0.2,0.5,0.8,1.0\}$.]{
    \includegraphics[width=0.3\linewidth]{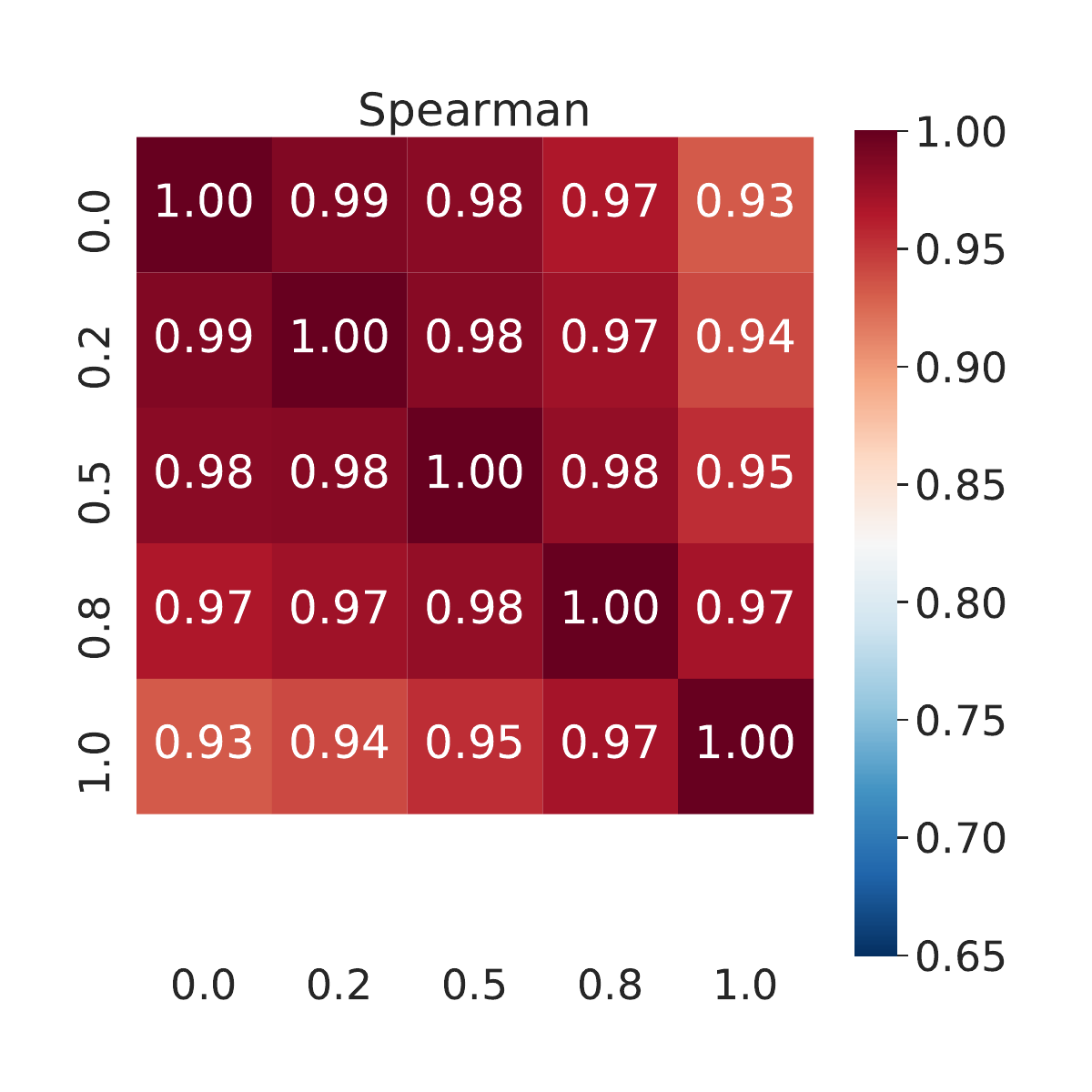}
    \includegraphics[width=0.3\linewidth]{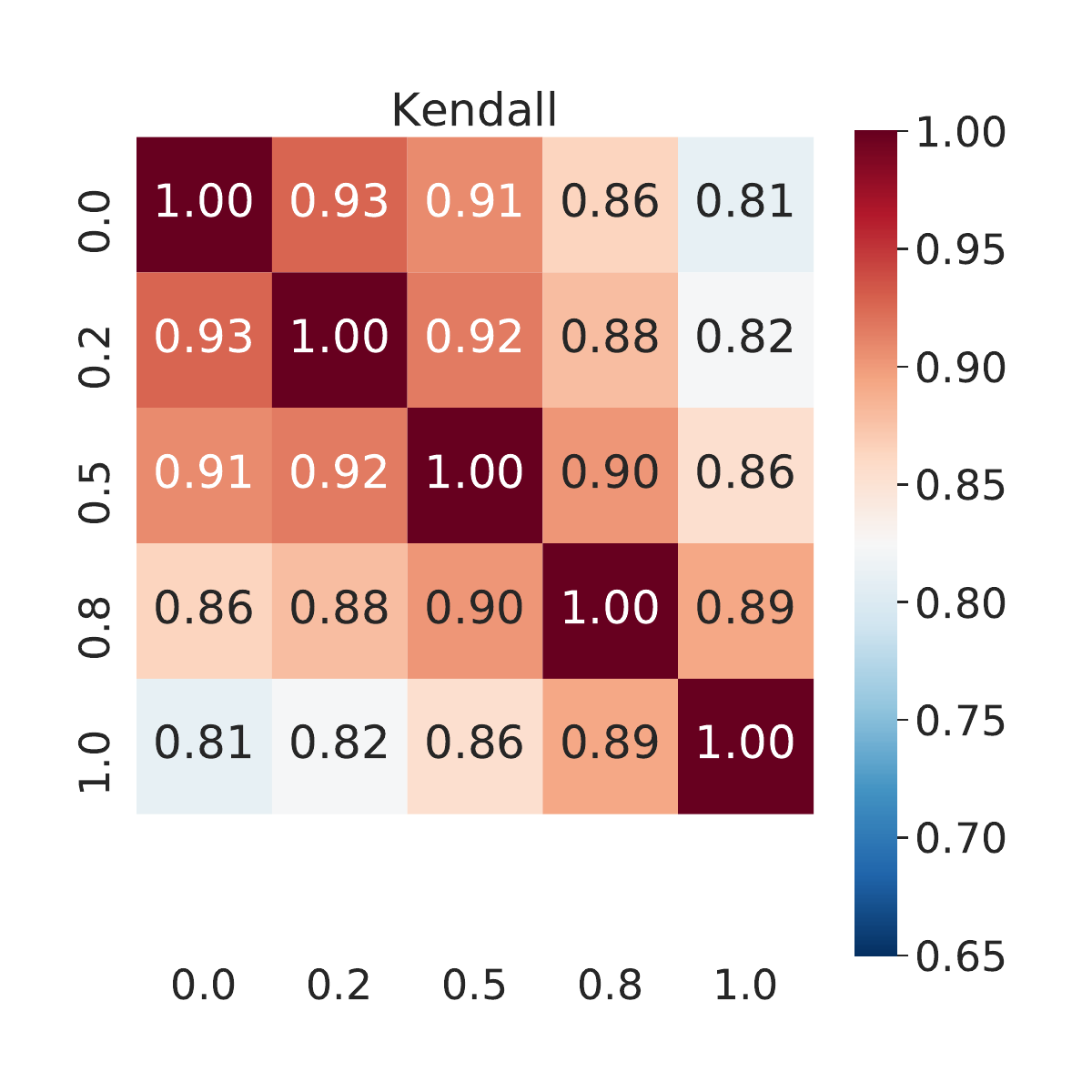}
    \includegraphics[width=0.3\linewidth]{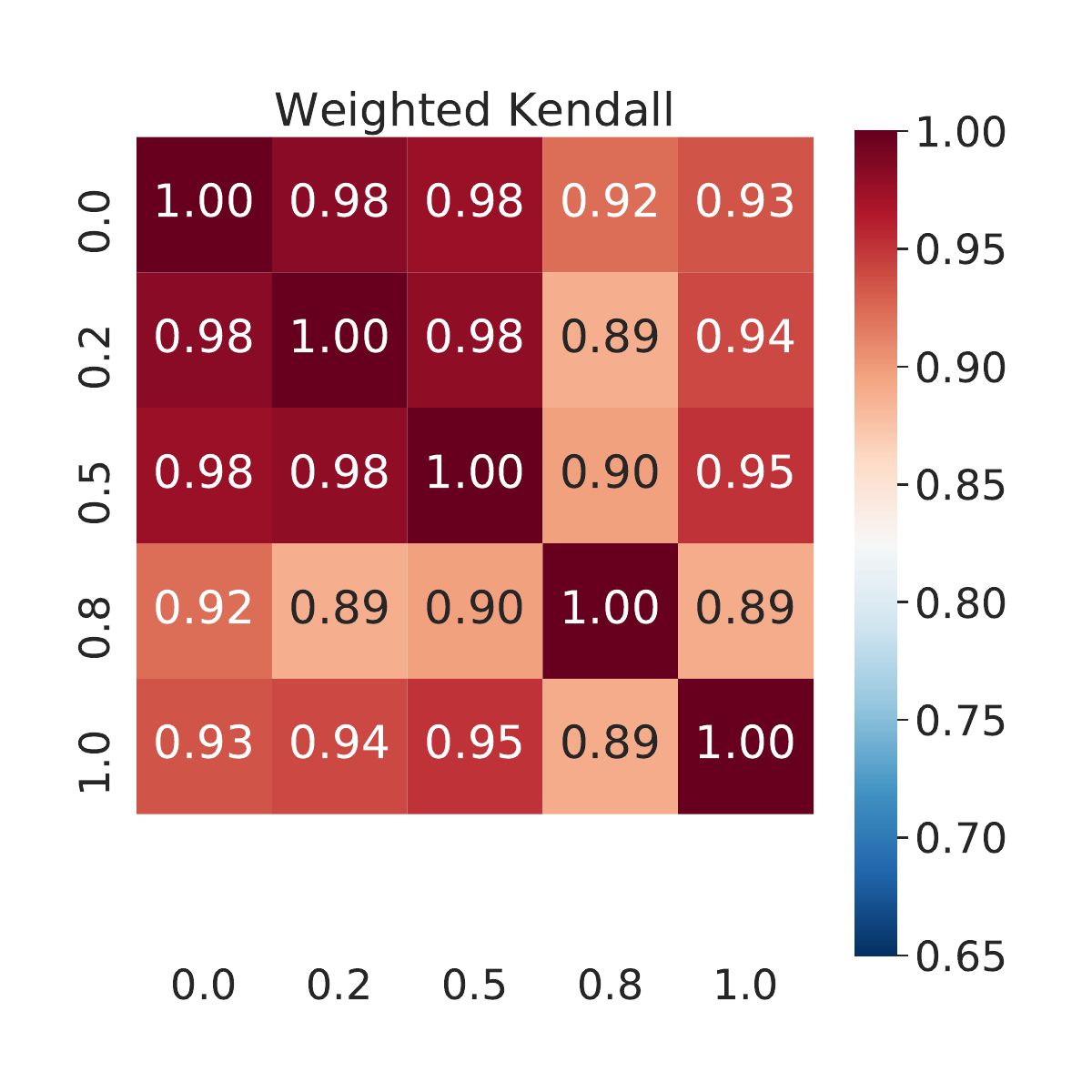}}
    \caption{Correlation of simulated daily node routing traffic \textbf{(top three)} and income \textbf{(bottom three)} with respect to different ratio of merchants among transaction endpoints $\epsilon$.
    }
    \label{fig:income_correlations}
\end{figure}

Our goal is to observe stable traffic characteristics throughout a sequence of days, measured as the correlation of node statistics across days. Towards this end, we measure the following node level summaries of the simulated traffic every day:
\begin{itemize}
    \item \emph{Routing traffic}: the number of transactions that are forwarded by a given node;
    \item \emph{Routing income}: the sum of all transaction fees that a given node charges for payment routing;
    \item \emph{Sender traffic}: the number of transactions that are initiated by a given node;
    \item \emph{Sender fee}: the sum of all transaction fees that a given node has to pay for his transactions to be forwarded by intermediary nodes.
\end{itemize}
In Figure~\ref{fig:income_correlations}, the Spearman, Kendall, unweighted and weighted Kendall-tau correlations of routing traffic and income are shown for $\epsilon=0.0$, 0.2, 0.5, 0.8, and 1.0.  For the definitions, see~\cite{vigna2015weighted}.

We observe high weighted Kendall-tau correlation, which means that the set of nodes with the highest routing income and traffic are very similar regardless of the ratio of merchants $\epsilon$ among transaction recipients. 

By contrast, we observe low values of (unweighted) Kendall-tau.  Since the set of nodes is dominated by low-traffic ones, the Kendall-tau value also depends mostly on the simulated traffic amount of these nodes. Hence, low Kendall-tau implies that nodes with low traffic and income fluctuate as transaction endpoints are selected at random. Most of these nodes have probably no traffic when transactions are centralized towards service providers ($\epsilon=1.0$). 

\begin{figure}[ht!]
    \centering
    \includegraphics[width=0.8\linewidth]{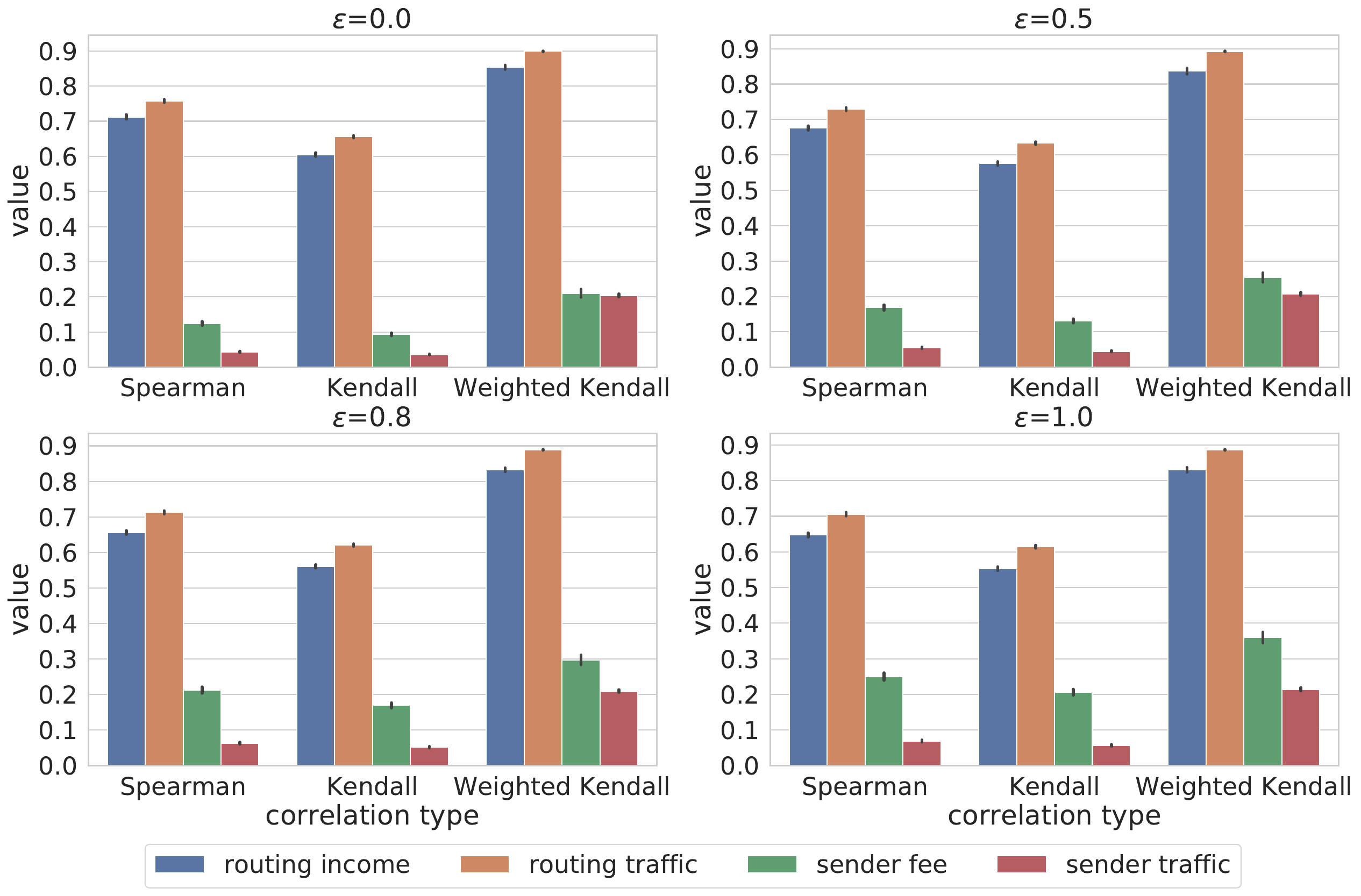}
    \caption{Mean Spearman, unweighted and weighted Kendall-tau cross correlation of node statistics over the $10$ independent simulations with respect to the ratio of merchants as transaction endpoints ($\epsilon\in\{0.0,0.5,0.8,1.0\}$).
    }
    \label{fig:meanCrossCorr}
\end{figure}

In Figure~\ref{fig:meanCrossCorr}, we assess the stability of the simulation by showing the mean correlation of four different node statistics over $10$ independent simulations for each snapshot.  Two of the statistics, routing income and routing traffic, show high correlation for all values of $\epsilon$, which means that nodes with high daily routing income and traffic are stable across independent experiments. By contrast, sender transaction fees and sender traffic especially vary highly, which is a natural consequence of uniform random sampling for source selection. By our measurements, ratio $\epsilon$ only affects the sender transaction fee. By increasing the value of $\epsilon$, more and more transactions are centralized towards merchants. Thus, sender nodes pay the transaction fees to more or less the same set of intermediary nodes, which results in higher sender transaction fee correlations.

\begin{figure}
\centering
\begin{minipage}{.4\textwidth}
    \centering
    \vspace{23pt}
    \includegraphics[width=\linewidth]{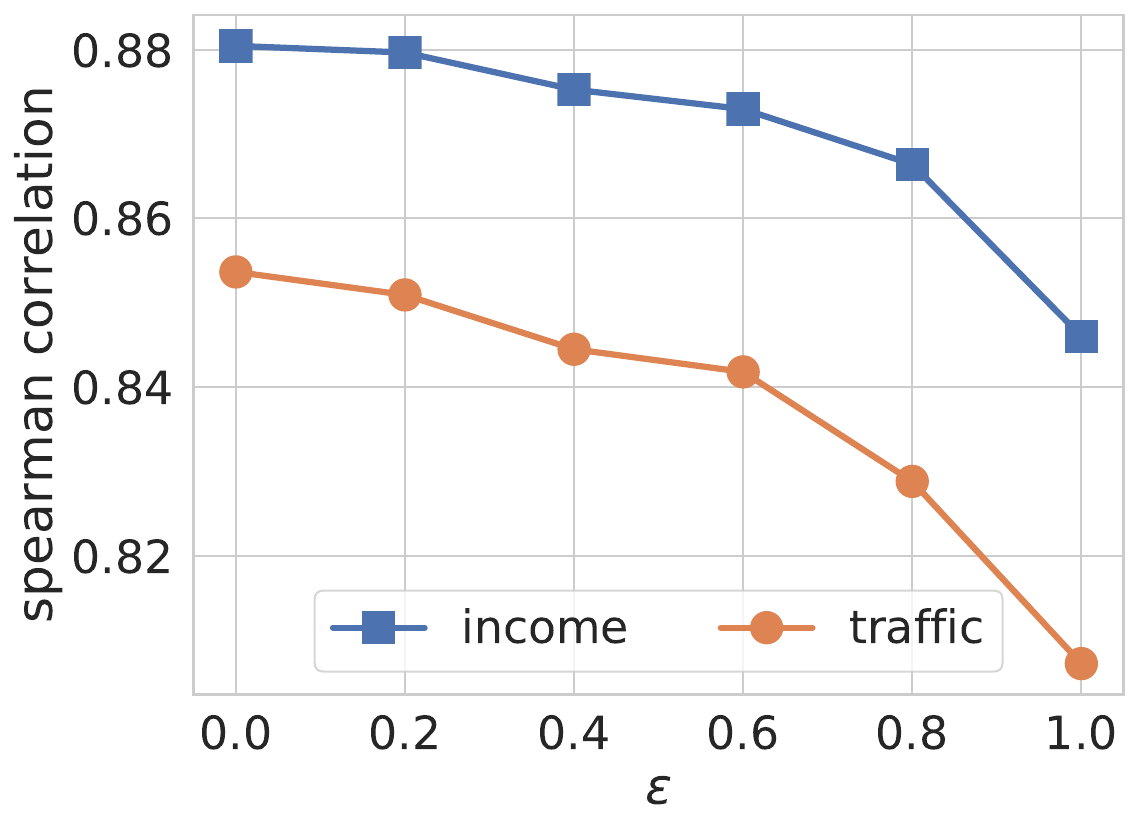}
    \caption{Spearman correlation of predicted daily routing income (or traffic) and  Betweeness centrality of LN nodes. The correlation decreases in case of high simulated merchant ratio $\epsilon$.}
    \label{fig:betweenness_correlations}
\end{minipage}
\hspace{6pt}
\begin{minipage}{.57\textwidth}
    \centering
    \includegraphics[width=\linewidth]{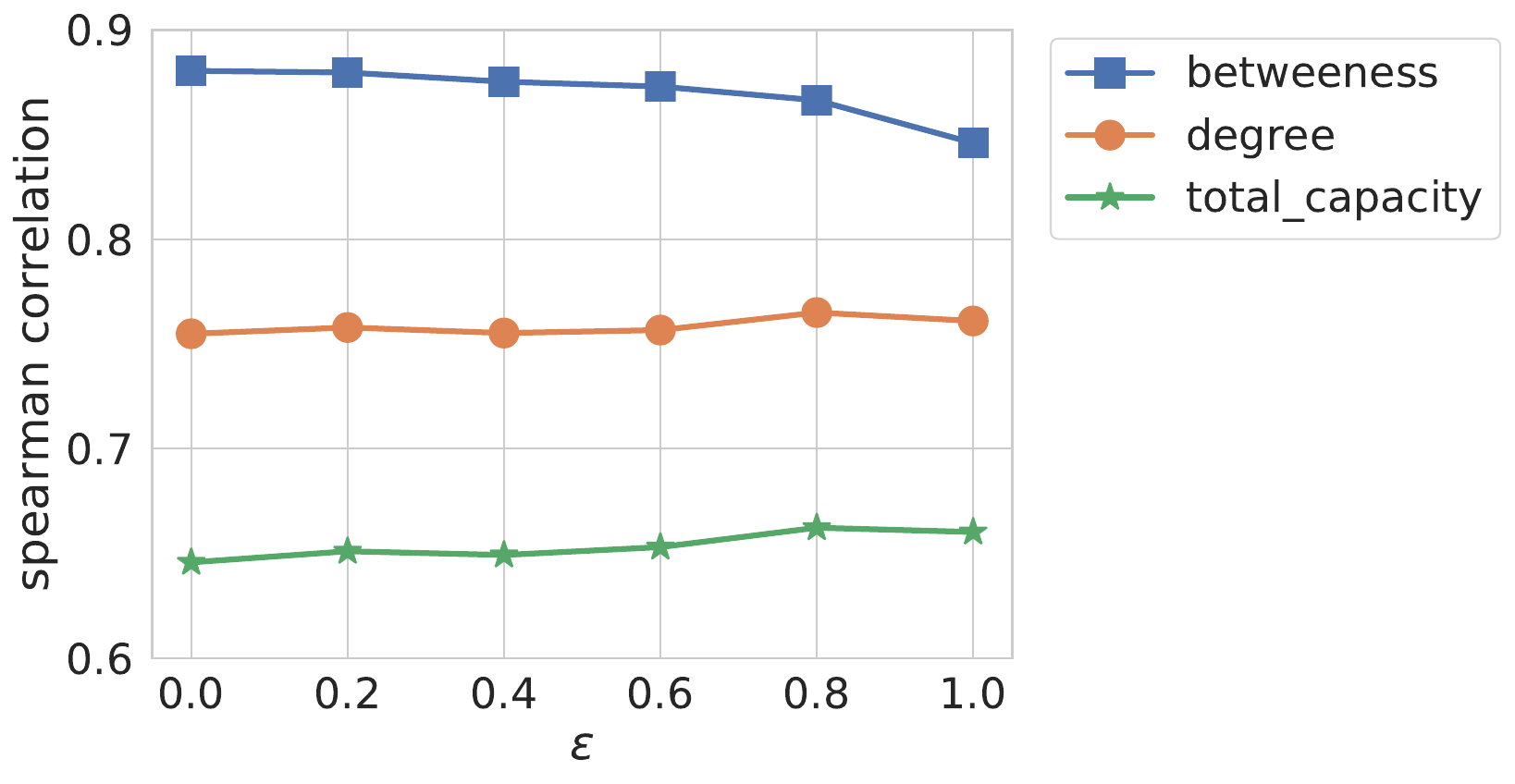}
    \caption{Spearman correlation of predicted daily routing income and graph centrality measures with regard to the merchant ratio $\epsilon$ among payment endpoints.}
    \label{fig:income_and_centrality}
\end{minipage}
\end{figure}

Finally, we compare our simulated routing income with simple estimates based on the properties of the nodes in LN as a graph. In a Youtube video, Pickhardt~\cite{earn2019pickhardt} shows the routing income of a node is proportional to its betweenness centrality in case the payment probability between any node pair is the same. In Figure~\ref{fig:betweenness_correlations}, we observe that our simulated routing income with parameters $\alpha=60,000$, $\tau=5000$, $\epsilon\in\{0.0,0.2,0.4,0.6,0.8,1.0\}$ is well correlated with the betweenness centrality of a node. However, the Spearman correlation decreases with larger $\epsilon$, which means  that since payment endpoints are biased towards merchants, we need a more accurate estimation method.
In Figure~\ref{fig:income_and_centrality}, we show two more node statistics, degree and total node capacity, both correlating much weaker to our prediction than betweenness centrality.

In summary, the set of nodes with high routing income and traffic are consistent across independent simulations regardless of the ratio of merchants among sampled transaction endpoints, while randomization naturally has a big influence on the low traffic end of the network. 
The low traffic end can be estimated by incorporating the role of a node in the simulation, as we do in a very simple way by controlling traffic towards merchants with the parameter $\epsilon$.

\section{Transaction Fee Competition} \label{sec:competition}

Our first analysis addresses the observed and potential profitability of LN, which is questioned in several blog posts~\cite{bitmex_post,lnbig_post}. 
A core value proposition of LN is that Bitcoin users can execute payments with negligible transaction fees. This feature may be cherished by payment initiators, but in case of insufficiently low network traffic, it could be unprofitable for router entities. 

Our goal is to assess how transaction costs depend on topology and to what extent they are targets to competition.
To measure transaction fee price competition, we use our traffic simulator to estimate daily node routing income and traffic volume for the $40$ consecutive LN snapshots in our data. 
Our findings on how revenue from routing depend on transaction fees shows a similar shape as experimented for BitMEX, a single LN node~\cite{bitmex_post}.

We use the parameters of the simulator that we calibrated based on published information on the income of certain nodes~\cite{lnbig_post} in Section~\ref{sec:param_tuning}.  
Our analysis in this section confirms that transaction fees are indeed very low, and they are potentially underpriced for relevant router nodes.

To analyze the competition that a node $x$ faces in the network, we compare the simulated traffic in a daily LN snapshot $G$ and in the graph $G_{x}$ that we obtain by removing node $x$ from $G$. By attempting to route the same set of $\tau$ transactions on $G$ and $G_x$, first of all we measure the number of failed payments $\varphi(x)$ that were originally routed through $x$ but are  incapable of reaching destination when $x$ is out of service. For each node $x$, the failure ratio of individual node traffic
 is $\frac{\varphi(x)}{\tau(x)}$ where $\tau(x)$ denotes the number of transactions through $x$ in the original simulation.

\begin{figure}
    \centering
    \includegraphics[width=0.35\linewidth]{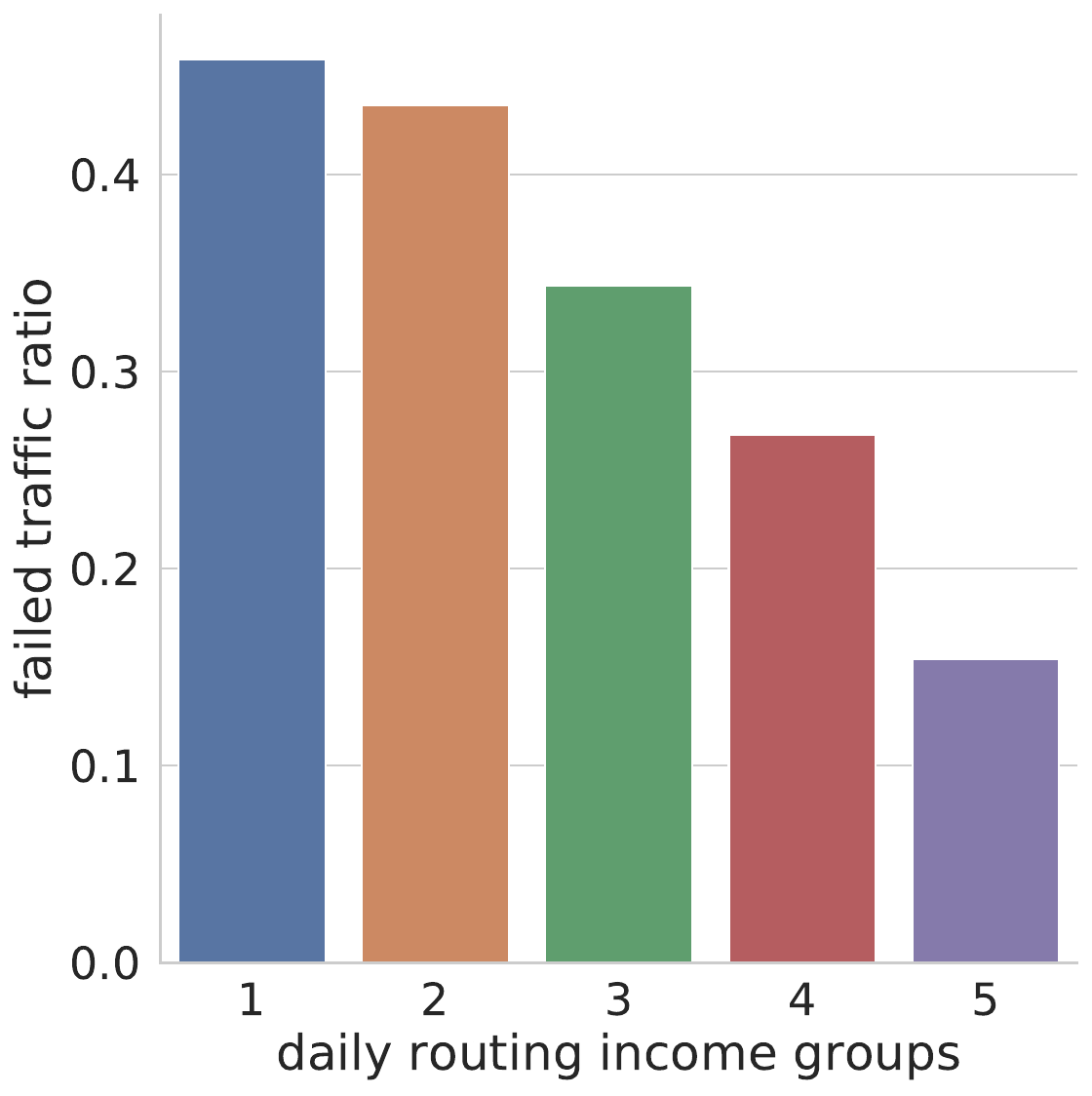}
    \caption{The average failure ratio of individual node traffic for five income groups defined as the top $1-10$, $11-20$, $21-50$, $50-100$, and $101-$ router nodes with highest simulated income.}
    \label{fig:failedTrafficRatio}
\end{figure}

In  Figure~\ref{fig:failedTrafficRatio}, we show the average ratio of the traffic of a node that has no alternate routing path, for five income groups defined as the top $1-10$, $11-20$, $21-50$, $50-100$, and $101-$ router nodes with highest simulated income.
For each group, the average is taken over its nodes $x$, considering the fraction of transactions $\frac{\varphi(x)}{\tau(x)}$ that cannot be routed anymore after removing $x$. It is interesting to observe that for the first three groups, the average ratio of traffic with no alternate path is at least $0.3$. This means that even if the 100 routers with highest simulated traffic increased their transaction fees close to on-chain fees, the majority of payment sources would have no less expensive option to route their payments.

\begin{figure}[ht!]
    \centering
    \includegraphics[width=0.35\linewidth]{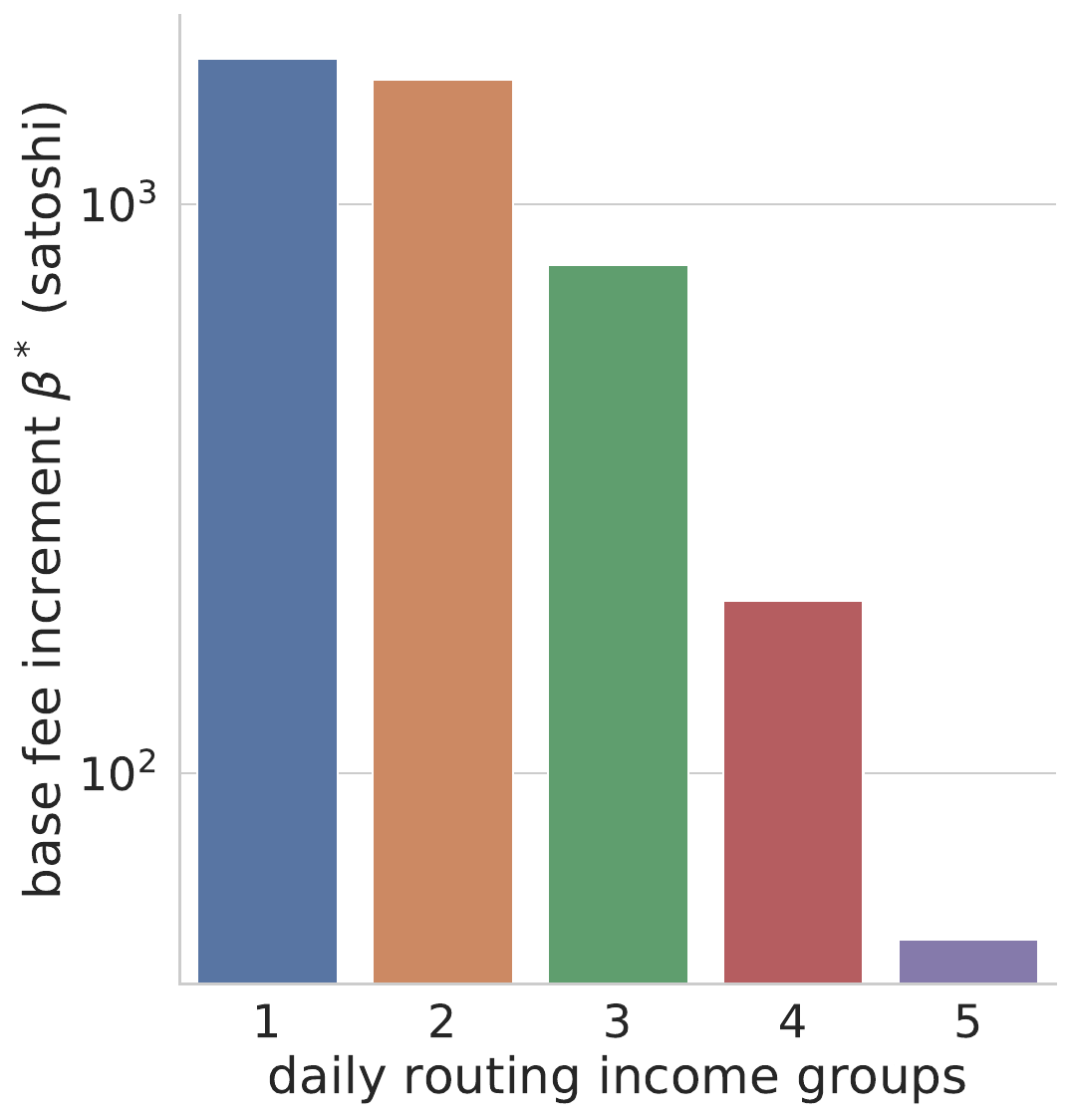}
    \hspace{12pt}
    \includegraphics[width=0.35\linewidth]{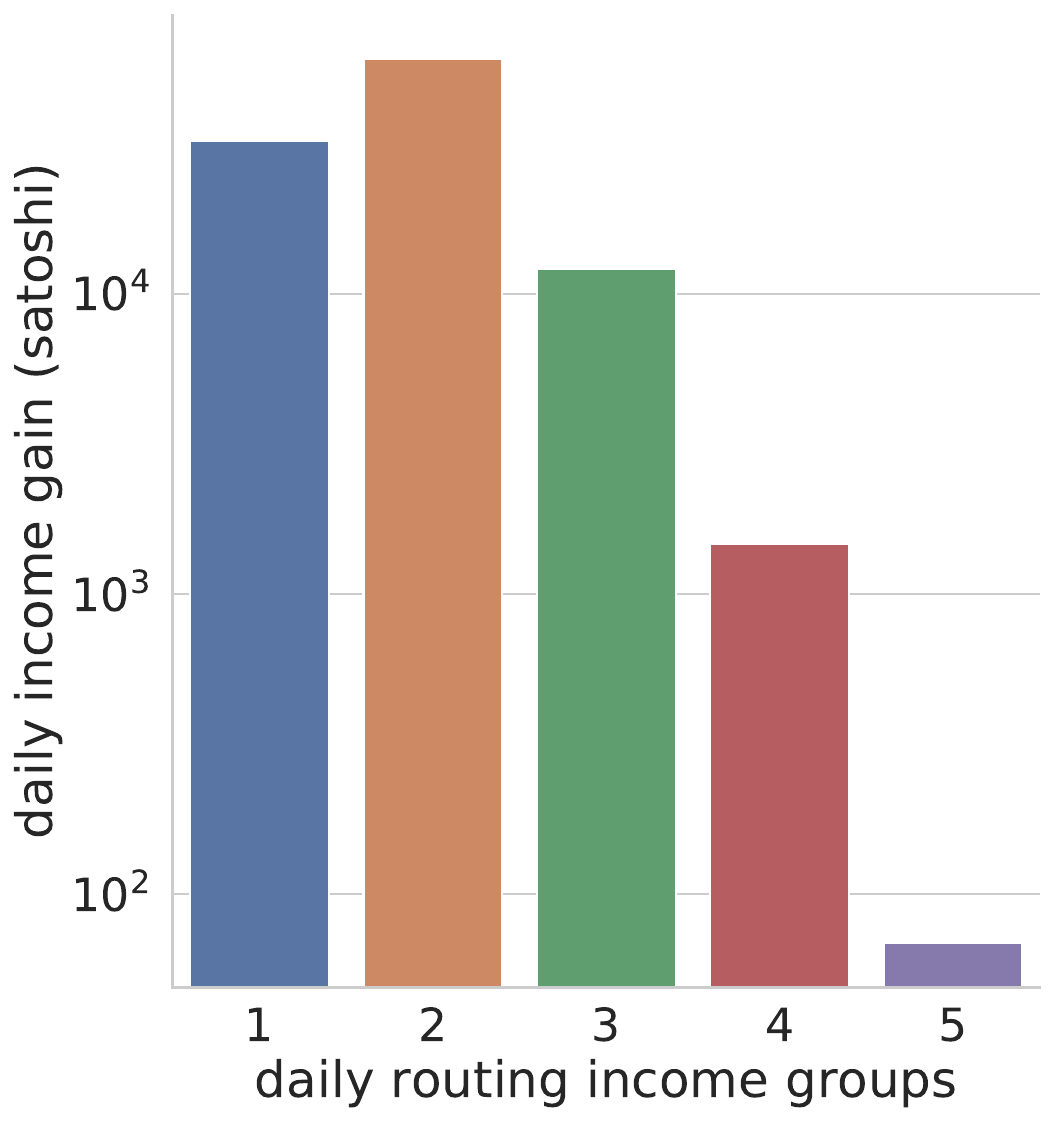}
    \caption{The maximal possible base fee increment ($\beta^*$, \textbf{left}), and the corresponding income gain \textbf{(right)} in satoshis, given the price competition assumptions in Section~\ref{sec:routing_fee}.
    Income groups are defined as the top $1-10$, $11-20$, $21-50$, $50-100$, and $101-$ router nodes with highest simulated income.}
    \label{fig:feeOptimalization}
\end{figure}

In the next experiment, we estimate the extent transaction prices are potentially limited by the competition among alternate routes in LN.
We take a highly pessimistic view by assuming that a transaction that can only be routed by relying on an intermediary node $x$ will select a payment method outside LN immediately if $x$ increases its transaction fees.  For other transactions, we search for the next cheapest route that avoids $x$ and assume that $x$ could increase its fees to match the second cheapest option. In other words, our analysis ignores the failed transactions $\varphi(x)$ and is based on the remaining $\tau(x)-\varphi(x)$ where payment routing avoiding node $x$ being available. For each of these transactions, the difference of the total fee $\delta$ can be calculated from the fees of the original path in $G$ and the alternative route in $G_x$. 

Our assumption is that if node $x$ increases its base fee by $\beta$, transactions with $\delta\geq\beta$ are still willing to pay for the additional costs, while for $\delta<\beta$, payments will be routed on the cheaper alternative path, where $\delta$ is the fee difference to the cheapest path avoiding $x$. Thus, by observing $\beta\geq 0$ at different thresholds, we propose an optimal $\beta^*$ base fee increment for each router node.

We estimate the optimal fee increase $\beta^*$ for each node over multiple snapshots and independent simulations. For the five node income groups that we previously defined in Figure~\ref{fig:failedTrafficRatio}, we show the average optimal base fee increment as well as the corresponding routing income gain in Figure~\ref{fig:feeOptimalization}. 

\begin{figure}
    \centering
    \includegraphics[width=0.3\linewidth]{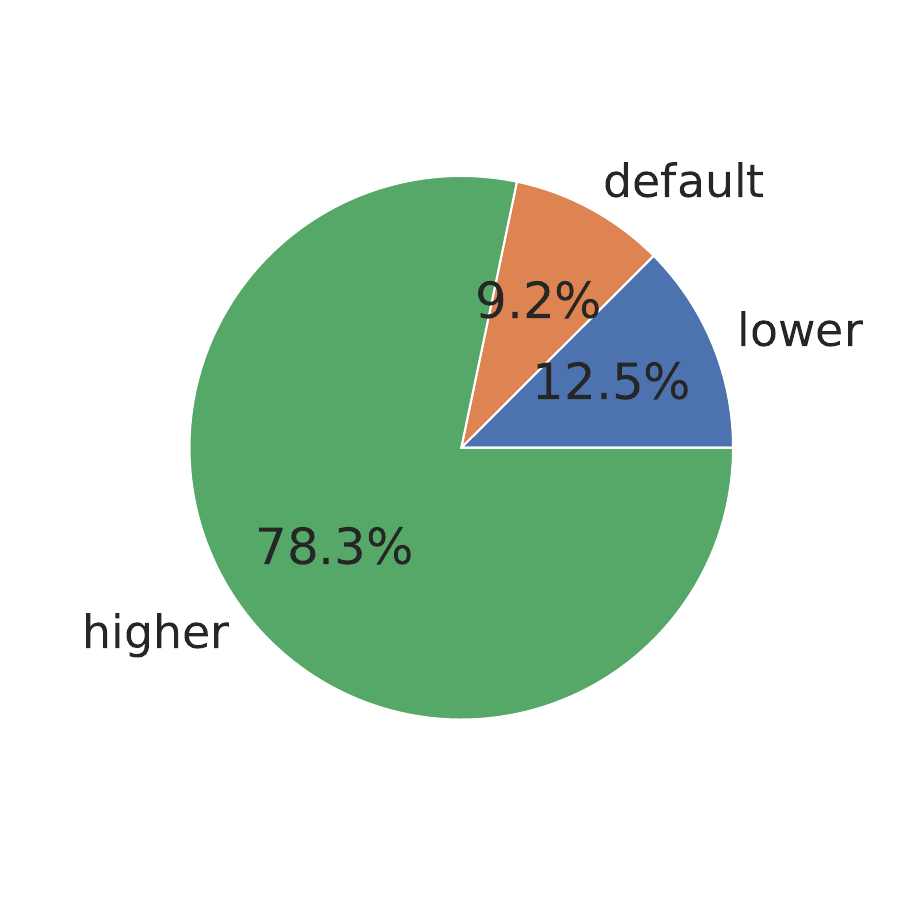}
    \includegraphics[width=0.3\linewidth]{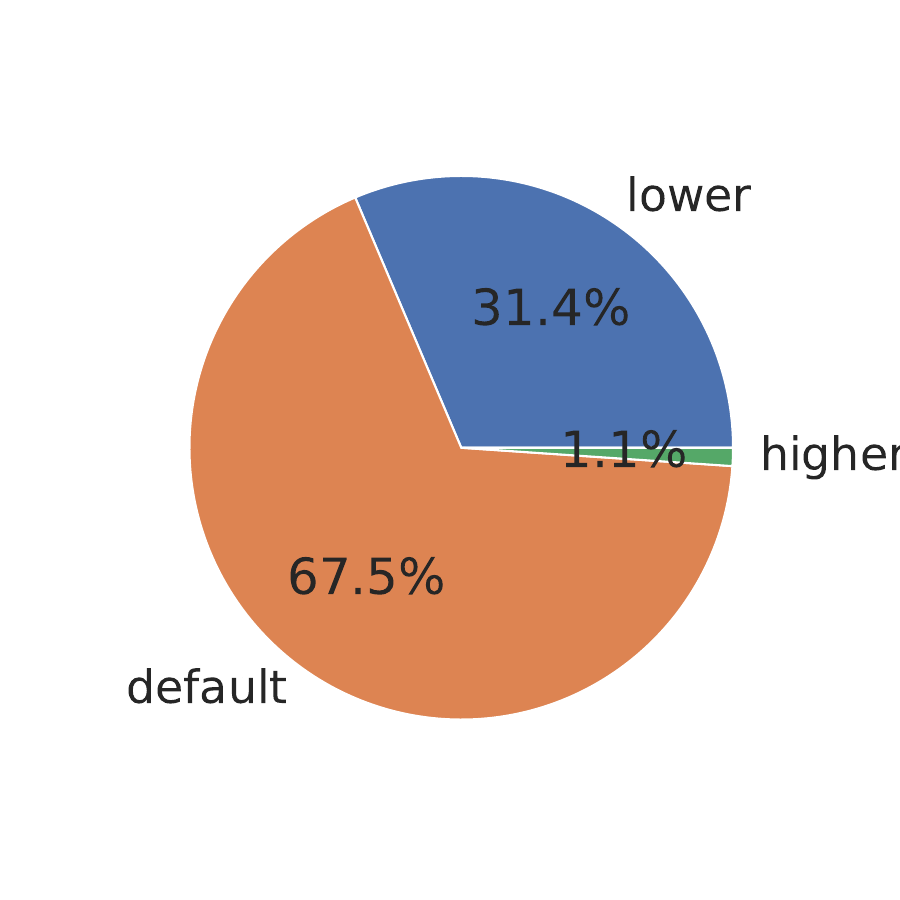}
    \includegraphics[width=0.3\linewidth]{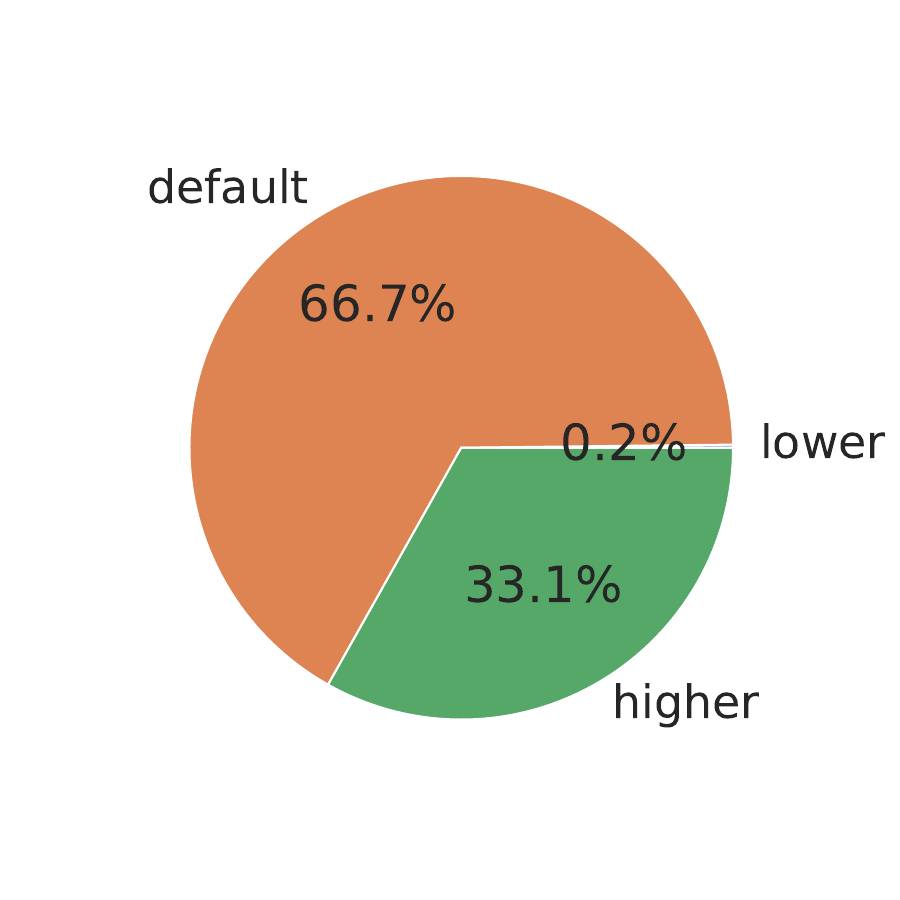}
    \caption{Distribution of channel capacities \textbf{(left)}, base fees \textbf{(center)} and fee rates \textbf{(right)} with regard to their default values in the \textit{lnd} client ($100000$ SAT, $1$ SAT, and $10^{-6}$ SAT), respectively.}
    \label{fig:channel_stat_normal}
\end{figure}

The transaction fee data shows that the current LN fee market is still immature, as  the majority of all channels apply default base fee ($1$ SAT) and fee rate ($10^{-6}$ SAT), while the capacities are usually set higher than the default value ($100000$ SAT) in the \textit{lnd} client, see Figure~\ref{fig:channel_stat_normal}.

In our measurements, we find that nodes with high routing income could still increase their base fee by a few hundred satoshis, thus generating an average gain of more than 10,000 satoshis in their daily income. Despite the low gain, our assumption is that it could get orders of magnitude higher if router nodes increased their base fee in succession, which could have a major impact on the competition for transaction costs.

\section{Profitability Estimation of Central Routers}
\label{sect:profitablity}

Router entities are an essential part of LN. They are responsible for keeping the network operational by forwarding payments. In this section, we estimate the current routing revenue of these central nodes, and give predictions how their income will change if the traffic over the current network increase. Note that our technique can also be used for node owners to predict the effect of opening and closing channels as well as changing capacities and transaction fees.

Central routing nodes are binding a huge amount of financial resources in the form of channel capacity, which enables them to serve high volumes of traffic. In general, router entities consist of a single node, but sometimes they have multiple LN nodes. For example, LNBIG.com owns $25$ nodes in our dataset. One of our main motivations was to estimate the annual return of investment (RoI) for entities by simulating daily traffic over several snapshots. In our measurements we calculate annual RoI as follows:
\begin{equation}
RoI = \frac{\text{estimated daily routing income in satoshis} \times 365}{\text{total amount of satoshis bound by channel capacities}}.
\label{eq:roi}    
\end{equation}

By simulating traffic with parameters $\tau=5,000$, $\alpha=60,000$, and $\epsilon=0.8$, we estimated the daily average income and traffic for each router. From these statistics and additional entity capacity data downloaded from 1ML.com, we estimate annual RoI in  Table~\ref{tab:roiTable}. We present all router entities with at least $50$ satoshis of simulated income and $10$ forwarded transactions per day on average. For each of these nodes, the following statistics are presented:
\begin{itemize}
    \item \emph{Entity capacity} as downloaded from 1ML.com. \emph{Capacity fraction} is the fraction of entity capacity and total network capacity. Remarkably, half of the total network capacity is bound by the nodes of LNBIG.com.
    \item Average \emph{transaction fee}, \emph{daily income}, and \emph{daily traffic}, based on the simulated mean cost in satoshis that a given entity charges for each payment routing over his channels during the observed $40$ snapshots, in ten random simulations, as explained in Section~\ref{sec:routing_fee}.
    \item \emph{Annual RoI} calculated from simulated daily income and entity capacity by Formula~\ref{eq:roi}. 
    \item \emph{Economical fee} in satoshis is the amount required on average to reach an annual 5\% RoI.  \emph{Fee ratio} is the ratio of the economical and the actual transaction fees. Higher values mean lower profitability.
    \item Three columns show the \emph{rank} of the nodes in decreasing order of annual RoI, total fee, and traffic.
\end{itemize}

Based on our findings, the annual RoI is way below $5$\% for almost all relevant entities. Only rompert.com achieved a comparable amount of annual RoI ($3.45\%$), who indeed applies orders of magnitude higher fees than others. It is interesting to see that despite its high transaction fees, it has the highest daily traffic in the simulation.  Note that rompert.com applies base fees close to onchain fees, which may invalidate the assumptions of our simulator if participants fall back to onchain rather than paying rompert.com routing fees.

Compared to the most profitable node rompert.com, the total estimated traffic of LNBIG.com through its $25$ nodes is only one third. The main reason behind low annual RoI is low transaction fees. Table~\ref{tab:roiTable} shows that for forwarding $\alpha=60,000$ satoshis, most of these entities ask for less then $100$ satoshis, which is less than $0.2$\% of the payment value. Very low fees may uphold LN's core value proposition, but they are economically irrational for the central routers holding the network together. Based  on our simulations, for several routers (e.g.,\ LNBIG.com, yalls.org, ln1.satoshilabs.com, etc.), fees should  be in the range of a few thousand satoshis to reach a $5$\% annual RoI, which is approximately the magnitude of on-chain transaction fees (1,000-2,000 satoshis\footnote{See \url{https://bitcoinfees.info/}.}).

\begin{table}
    \centering
    \Rotatebox{90}{%
    \begin{tabular}{lrrrrrrrrrrrrr}
\toprule
                           &  Capacity &      Capacity &   Fee for &  Daily &  Daily         &  Annual   &  Fee                & Economical       &  RoI              &     Fee         &  Traffic \\
                    name   &fraction (\%)&  (millions) &   60000SAT &  income&  traffic       &  RoI (\%) &  ratio              & fee (SAT)             &  rank             & rank            &  rank \\
\midrule
             rompert.com &      0.958 &      969 &     4371.9 &       91831.4 &          835.2 &    3.458924 &                 1.4 &            6319.7 &                1 &               1 &                   1 \\
               zigzag.io &      0.926 &      950 &      601.0 &        6716.2 &           11.2 &    0.258036 &                19.4 &           11645.7 &                2 &               2 &                  23 \\
               LNBIG.com &     52.309 &    53686 &       32.4 &        5932.0 &          288.2 &    0.004033 &              1239.8 &           40181.0 &               17 &               8 &                   4 \\
               yalls.org &      1.728 &     1772 &      151.0 &        2001.2 &           13.3 &    0.041199 &               121.4 &           18325.8 &                4 &               4 &                  22 \\
     ln1.satoshilabs.com &      2.597 &     2665 &       60.0 &        1180.5 &           19.7 &    0.016167 &               309.3 &           18556.9 &                9 &               6 &                  20 \\
                OpenNode &      1.364 &     1400 &       87.5 &         825.6 &           29.6 &    0.021519 &               232.4 &           20335.2 &                6 &               5 &                  17 \\
               tippin.me &      1.054 &     1081 &       55.1 &         474.4 &           45.3 &    0.016009 &               312.3 &           17207.9 &               10 &               7 &                  12 \\
              BlueWallet &      1.523 &     1562 &      276.0 &         453.4 &           16.4 &    0.010588 &               472.2 &          130329.2 &               13 &               3 &                  21 \\
 LightningPowerUsers.com &      2.413 &     2440 &        1.4 &         368.7 &          305.1 &    0.005514 &               906.7 &            1278.1 &               16 &              19 &                   3 \\
                   ACINQ &      3.367 &     3455 &        7.0 &         281.7 &           40.2 &    0.002975 &              1680.5 &           11763.3 &               18 &              13 &                  14 \\
      btc.lnetwork.tokyo &      0.240 &      245 &        2.8 &         224.0 &          107.6 &    0.033247 &               150.4 &             426.8 &                5 &              16 &                   7 \\
           Sagittarius A &      0.528 &      541 &       14.9 &         218.3 &           30.9 &    0.014715 &               339.8 &            5047.4 &               11 &              11 &                  16 \\
      1ML.com node ALPHA &      0.688 &      706 &        1.1 &         174.5 &          164.6 &    0.009021 &               554.3 &             587.5 &               15 &              20 &                   5 \\
        tady je slushovo &      0.413 &      423 &        1.1 &         123.2 &          116.3 &    0.010622 &               470.7 &             499.0 &               12 &              21 &                   6 \\
   Electrophorus [W\_C\_B] &      0.384 &      393 &       19.5 &         101.7 &           80.8 &    0.009425 &               530.5 &           10346.2 &               14 &              10 &                   9 \\
   There be dragons here &      0.172 &      176 &       25.5 &          87.5 &            7.1 &    0.018050 &               277.0 &            7068.2 &                7 &               9 &                  25 \\
          LightningTo.Me &      1.202 &     1233 &        0.4 &          78.3 &          725.5 &    0.002317 &              2158.3 &             919.1 &               20 &              25 &                   2 \\
            fairly.cheap &      1.365 &     1401 &        0.7 &          74.9 &          103.0 &    0.001950 &              2563.7 &            1863.8 &               24 &              24 &                   8 \\
           Bitrefill.com &      2.323 &     2383 &        4.4 &          73.2 &           31.4 &    0.001121 &              4460.8 &           19713.9 &               25 &              15 &                  15 \\
          BOLTENING.club &      0.898 &      921 &        6.7 &          54.7 &           43.3 &    0.002166 &              2308.8 &           15362.1 &               22 &              14 &                  13 \\
          ORANGESQUIRREL &      0.018 &       18 &        7.0 &          52.7 &            7.5 &    0.104882 &                47.7 &             333.8 &                3 &              12 &                  24 \\
  lightning-roulette.com &      0.728 &      747 &        1.1 &          49.6 &           46.8 &    0.002421 &              2065.7 &            2189.6 &               19 &              23 &                  10 \\
                CoinGate &      0.821 &      842 &        1.1 &          49.2 &           46.4 &    0.002129 &              2348.6 &            2489.5 &               23 &              22 &                  11 \\
       ln.BitSoapBox.com &      0.590 &      605 &        1.6 &          37.8 &           23.9 &    0.002276 &              2196.6 &            3503.3 &               21 &              18 &                  18 \\
       Blockstream Store &      0.076 &       78 &        1.6 &          37.1 &           23.2 &    0.017364 &               287.9 &             460.7 &                8 &              17 &                  19 \\
\bottomrule
\end{tabular}

}%
    \caption{Estimated daily income, traffic and annual RoI for relevant router entities.  Columns are explained in Section~\ref{sect:profitablity}.
    Note that currently on-chain transaction fees for a regular transaction (2 inputs, 2 outputs) is in the range of 1000-2000 satoshis.}
    \label{tab:roiTable}
\end{table}

\begin{figure}
    \centering
    \includegraphics[width=0.8\linewidth]{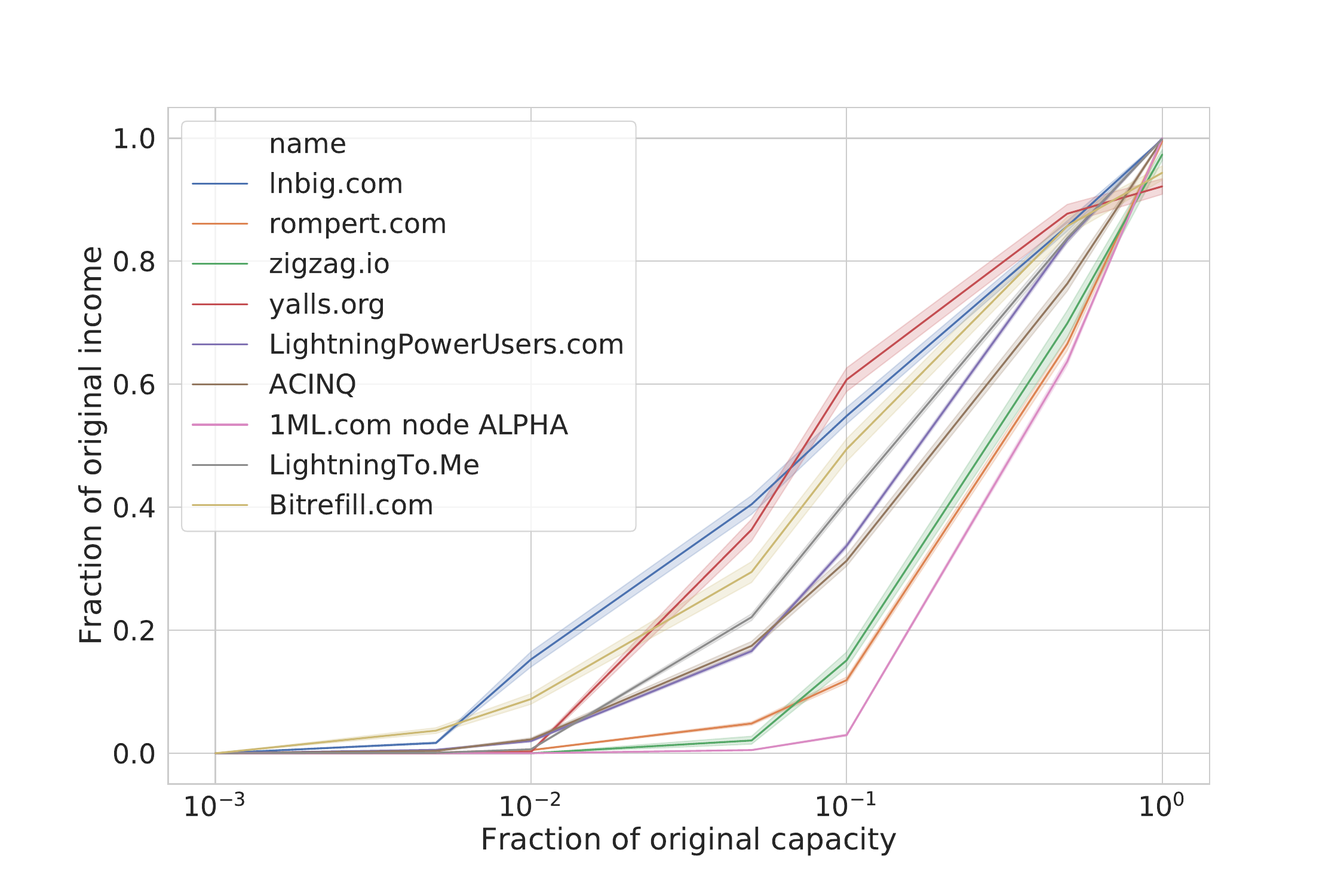}
    \caption{The remaining fraction of the original estimated daily routing income, after reducing node capacities to the given fractions.}
    \label{fig:income_frac}
\end{figure}

\begin{figure}
    \centering
    \includegraphics[width=0.8\linewidth]{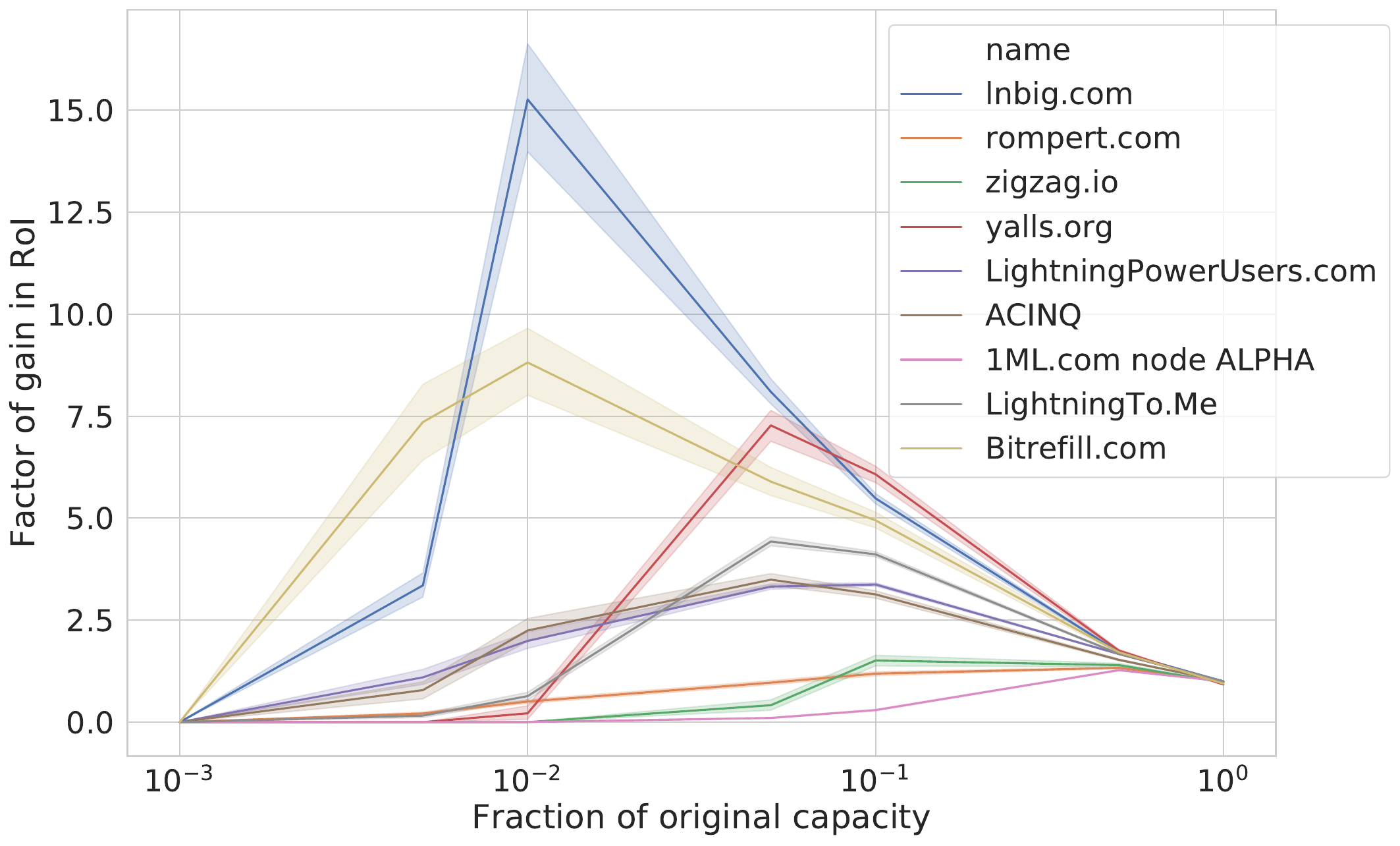}
    \caption{RoI gain after reducing node capacities to the given fractions.}
    \label{fig:roi_gain}
\end{figure}

Capacity overprovisioning also causes low RoI. For example, extremely large LNBIG.com capacities result in low RoI, despite the reasonable daily income reported.  By using our traffic simulator, we observed that the router entities of Table~\ref{tab:roiTable} can increase their RoI by reducing their channel capacities.  For each of these routers, we estimated the changes in  revenue (Figure~\ref{fig:income_frac}) and RoI (Figure~\ref{fig:roi_gain}), after reducing all of its edge capacities to $50, 10, 5, 1, 0.5, 0.1$\% of the original value, with the assumption that all other routers keep their capacities. In our measurements, LNBIG.com can  significantly improve its RoI by bounding only $1$\% of its original capacity values. In Table~\ref{tab:roiOpt}, we compute the estimated optimal RoI for the central routers.

\begin{table}
    \centering

\begin{tabular}{l|rrrrrrr}
\toprule
                    Entity &  RoI gain &  Capacity &  Income &  Original &    Optimal &  Optimal &  Original \\
                    name &  (times) &  fraction &  fraction &  RoI (\%) &    RoI (\%) &  RoI rank &  RoI rank \\
\midrule
               lnbig.com &   15.263039 &      0.01 &     0.152630 &   0.004033 &  0.061557 &           5.0 &            17.0 \\
           Bitrefill.com &    8.815776 &      0.01 &     0.088158 &   0.001121 &  0.009881 &          21.0 &            25.0 \\
               yalls.org &    7.274128 &      0.05 &     0.363706 &   0.041199 &  0.299685 &           3.0 &             4.0 \\
            fairly.cheap &    5.527895 &      0.01 &     0.055279 &   0.001950 &  0.010781 &          18.0 &            24.0 \\
          LightningTo.Me &    4.428039 &      0.05 &     0.221402 &   0.002317 &  0.010258 &          20.0 &            20.0 \\
       ln.BitSoapBox.com &    4.270262 &      0.10 &     0.427026 &   0.002276 &  0.009720 &          22.0 &            21.0 \\
                   ACINQ &    3.492428 &      0.05 &     0.174621 &   0.002975 &  0.010391 &          19.0 &            18.0 \\
 LightningPowerUsers.com &    3.374553 &      0.10 &     0.337455 &   0.005514 &  0.018608 &          13.0 &            16.0 \\
       Blockstream Store &    3.211826 &      0.10 &     0.321183 &   0.017364 &  0.055771 &           6.0 &             8.0 \\
     ln1.satoshilabs.com &    3.165573 &      0.05 &     0.158279 &   0.016167 &  0.051176 &           8.0 &             9.0 \\
   There be dragons here &    3.064046 &      0.05 &     0.153202 &   0.018050 &  0.055307 &           7.0 &             7.0 \\
   Electrophorus [W\_C\_B] &    2.229242 &      0.10 &     0.222924 &   0.009425 &  0.021010 &          11.0 &            14.0 \\
  lightning-roulette.com &    1.724278 &      0.10 &     0.172428 &   0.002421 &  0.004174 &          23.0 &            19.0 \\
              BlueWallet &    1.672075 &      0.05 &     0.083604 &   0.010588 &  0.017704 &          15.0 &            13.0 \\
               zigzag.io &    1.511212 &      0.10 &     0.151121 &   0.258036 &  0.389947 &           2.0 &             2.0 \\
                OpenNode &    1.458822 &      0.50 &     0.729411 &   0.021519 &  0.031392 &          10.0 &             6.0 \\
        tady je slushovo &    1.445600 &      0.50 &     0.722800 &   0.010622 &  0.015355 &          16.0 &            12.0 \\
          BOLTENING.club &    1.430646 &      0.10 &     0.143065 &   0.002166 &  0.003098 &          24.0 &            22.0 \\
      btc.lnetwork.tokyo &    1.422923 &      0.50 &     0.711462 &   0.033247 &  0.047307 &           9.0 &             5.0 \\
                CoinGate &    1.400418 &      0.50 &     0.700209 &   0.002129 &  0.002981 &          25.0 &            23.0 \\
             rompert.com &    1.330968 &      0.50 &     0.665484 &   3.458924 &  4.603716 &           1.0 &             1.0 \\
          ORANGESQUIRREL &    1.313521 &      0.50 &     0.656760 &   0.104882 &  0.137764 &           4.0 &             3.0 \\
               tippin.me &    1.297128 &      0.50 &     0.648564 &   0.016009 &  0.020766 &          12.0 &            10.0 \\
      1ML.com node ALPHA &    1.273918 &      0.50 &     0.636959 &   0.009021 &  0.011491 &          17.0 &            15.0 \\
           Sagittarius A &    1.216862 &      0.50 &     0.608431 &   0.014715 &  0.017906 &          14.0 &            11.0 \\
\bottomrule
\end{tabular}

\caption{Estimated optimal channel capacity reduction for maximal RoI of the routers of Table~\ref{tab:roiTable}. Capacity fraction is the estimated optimal fraction of the original channel capacities and income fraction is the estimated fraction of the original income by using reduced channel capacities.}
    \label{tab:roiOpt}
\end{table}

To estimate whether routers can be more profitable with an increase in traffic volume or transaction values, we ran simulations with different values of $\tau$ and $\alpha$ and measured the fraction of unsuccessful payments as well as the average length of completed payment paths.

\begin{figure}
    \centering
    \includegraphics[width=\linewidth]{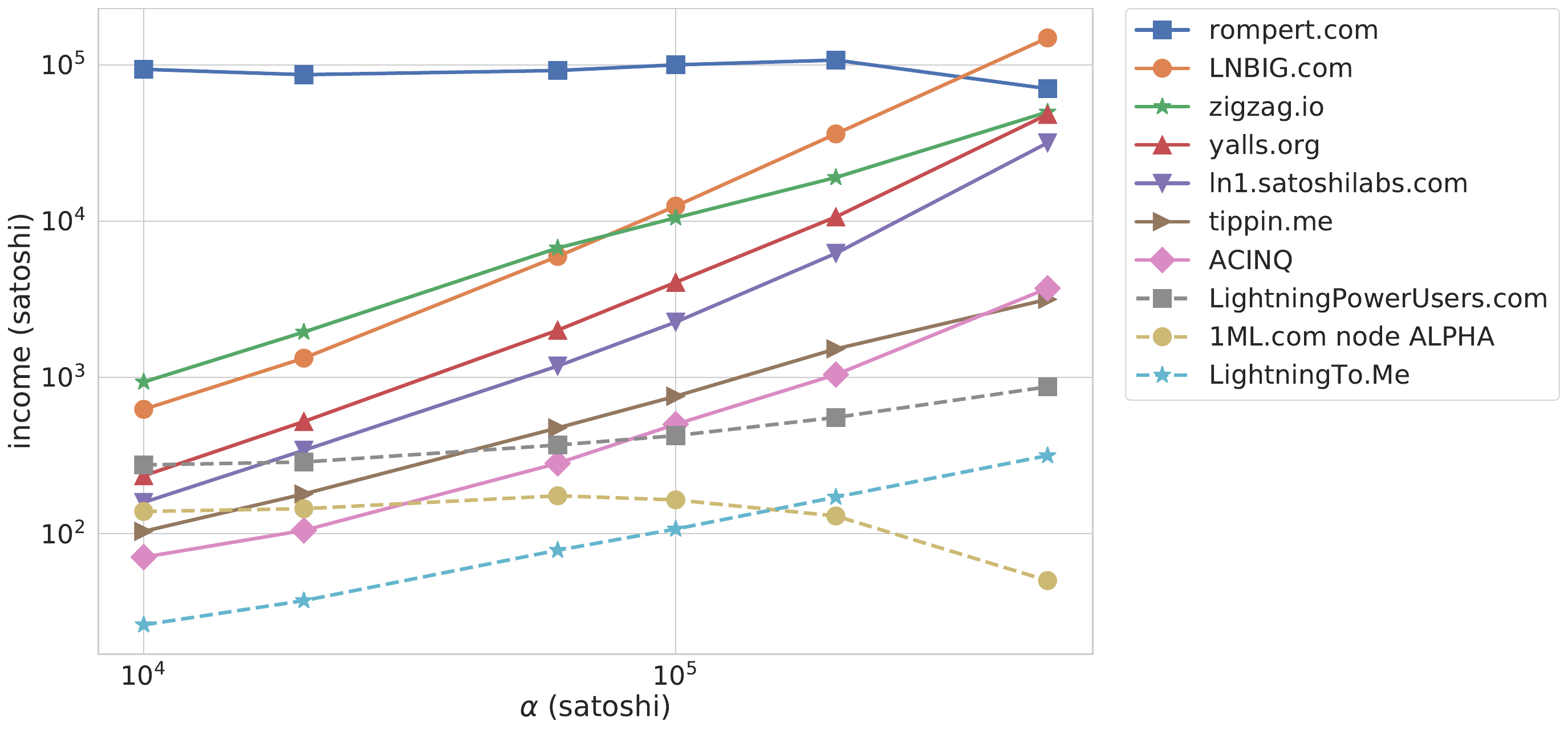}
    \caption{Average simulated daily routing income of some LN router entities as the function of the transaction value $\alpha$.}
    \label{fig:avgDailyIncome}
\end{figure}

\begin{figure}
    \centering
    \includegraphics[width=\linewidth]{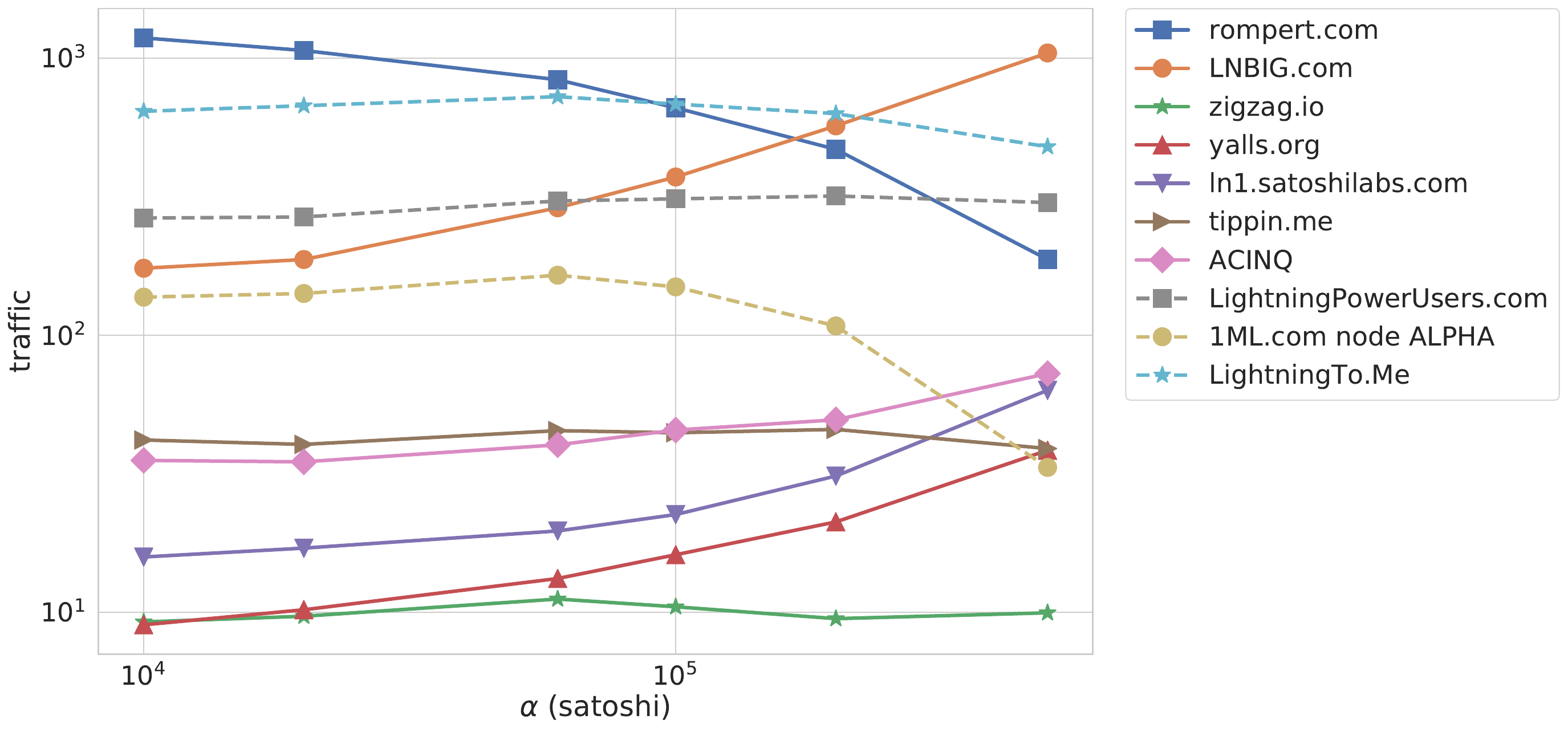}
    \caption{Average simulated daily routing traffic of some LN router entities as the function of the transaction value $\alpha$.}
    \label{fig:avgDailyTraffic}
\end{figure}

First we vary the transaction value $\alpha$ with a fixed number of daily transactions $\tau=5,000$. In Figures~\ref{fig:avgDailyIncome} and~\ref{fig:avgDailyTraffic}, we present statistics for ten central entities based on their service profiles. For example, zigzag.io is a cryptocurrency exchange service, while ACINQ provides solutions for Bitcoin scalability. Additional entity profiles can be found in Table~\ref{tab:profile}. In Figure~\ref{fig:avgDailyIncome}, the income for most of the nodes significantly increases with transaction value, while this effect is almost negligible for rompert.com, LightningPowerUsers.com, and 1ML.com node ALPHA, whose behavior can be explained by charging almost only a base fee and applying a fee rate close to zero.

\begin{figure}
    \centering
    \includegraphics[width=0.8\linewidth]{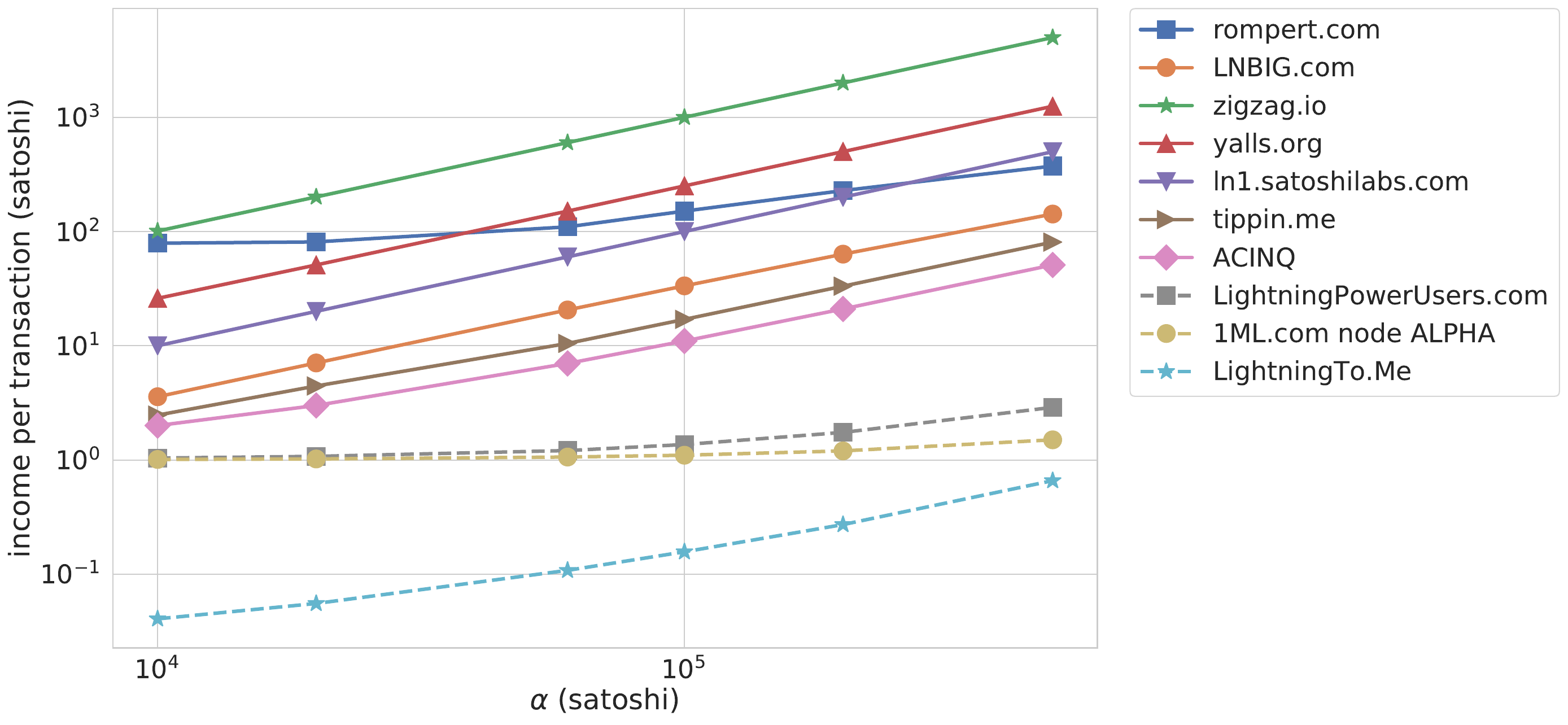}
    \caption{Average simulated daily routing income per transaction for some LN router entities as the function of the transaction value $\alpha$.}
    \label{fig:avgDailyEfficiency}
\end{figure}

The simulated amount of daily traffic for the ten central nodes is shown in  Figure~\ref{fig:avgDailyTraffic}. We observe that scalability and capacity providers LightningTo.Me, LightningPowerUsers.com, and 1ML.com node ALPHA are responsible for forwarding a significant amount of payments irrespective of $\alpha$. Probably due to the lack of high capacity channels, the traffic of rompert.com and 1ML.com node ALPHA drop at $\alpha=500,000$ satoshis ($\approx 41$ USD). By contrast, the number of payments routed by LNBIG.com increases with payment value due to the fact that this entity owns approximately half of all network capacity, as seen Table~\ref{tab:roiTable}. In Figure~\ref{fig:avgDailyEfficiency}, we provide an efficiency metric for each entity by dividing estimated income by traffic volume. The efficiency of rompert.com and LNBIG.com are surpassed by zigzag.io and yalls.org for $\alpha\geq60,000$ satoshis, as these service providers have reasonable routing income relative to the number of daily forwarded transactions. On the other hand, LightningPowerUsers.com, 1ML.com node ALPHA, and LightningTo.Me have orders of magnitude lower efficiency than other relevant entities. They are likely not considering routing profitability, as their transaction fees are negligible.

\begin{table}
\begin{tabular}{|l|l|}
\hline
Entity name                & Service profile   \\
\hline
rompert.com             & Provider of  some Lightning Network related information \\
LNBIG.com               & Half of the total network capacity in bound by the nodes of this entity \\
zigzag.io               & Exchange Top Cryptocurrencies in seconds with low fees \\
yalls.org               & Read and write articles, with Lightning Network micropayments.  \\
ln1.satoshilabs.com     & Cryptocurrency solution developers \\
tippin.me               & Send and receive Bitcoin tips on Twitter  \\
ACINQ                   & One of the leading companies working on Bitcoin scalability   \\
1ML.com node ALPHA      & Lightning Network Search and Analysis Engine  \\
LightningTo.Me          & Helping to resolve routing and capacity issues  \\
LightningPowerUsers.com & Request Inbound Capacity  \\
\hline
\end{tabular}
\caption{LN network entities with related service profiles.}
\label{tab:profile}
\end{table}

\begin{figure}
    \centering
    \includegraphics[width=0.9\linewidth]{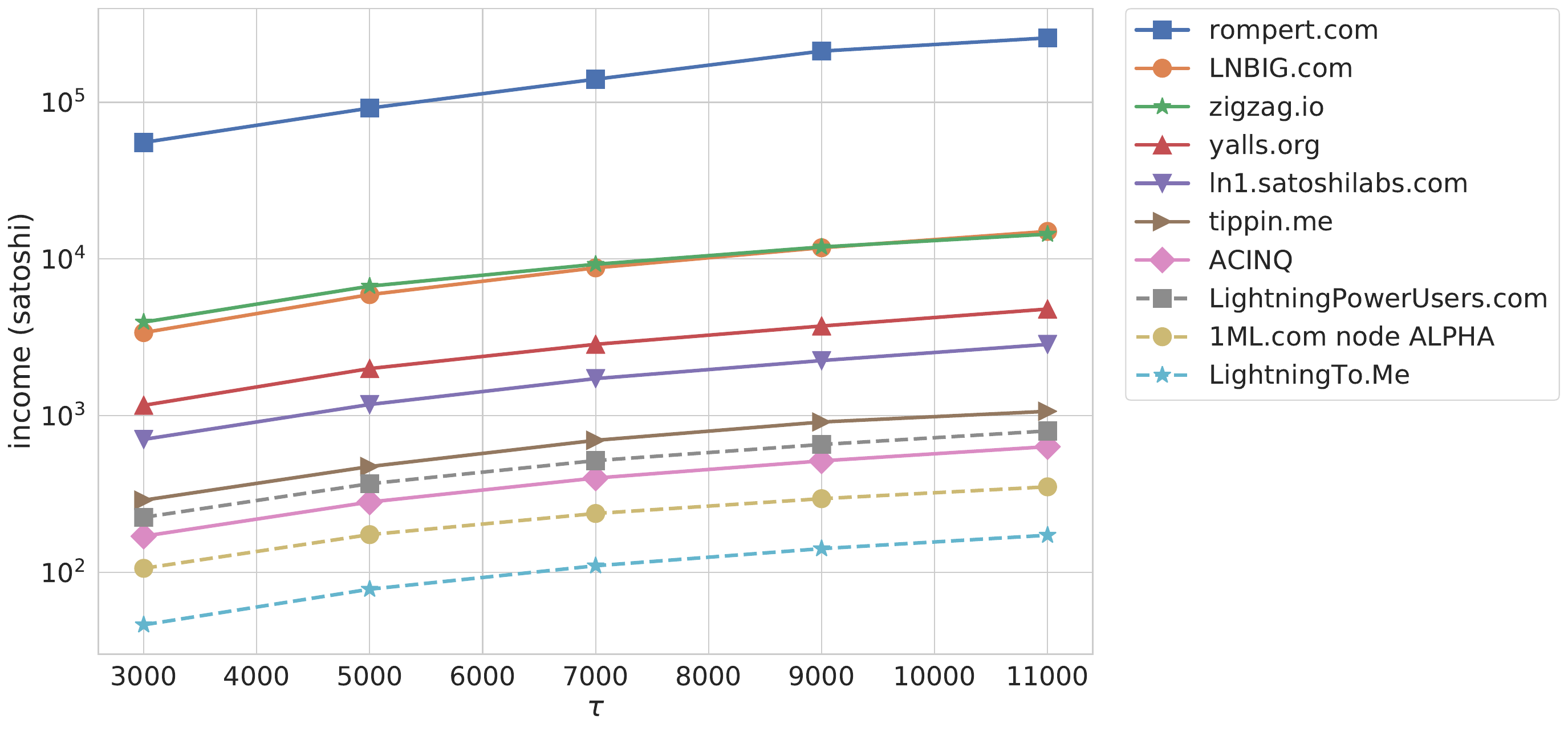}
    \includegraphics[width=0.9\linewidth]{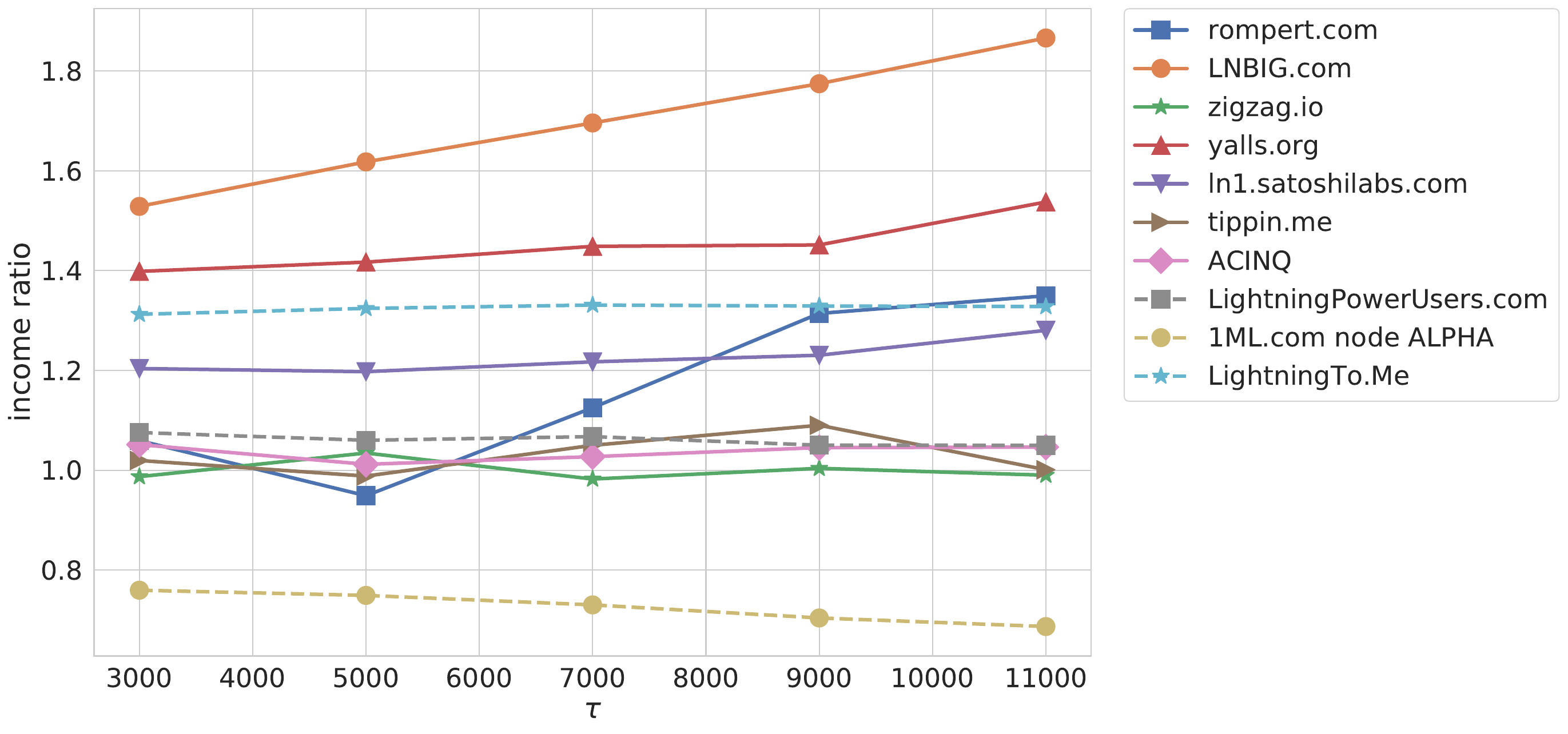}
    \caption{Average simulated daily routing income \textbf{(top)} and the income divided by the optimistic income when channel depletion is ignored \textbf{(bottom)} for some LN router entities as the function of the simulated transaction count $\tau$.  Note that the ratio is above 1 for most nodes as they can take over routing for depleted channels.}
    \label{fig:income-tau}
\end{figure}

Next, we estimate the effect of channel depletion, which can be a side-effect of increasing the traffic without increasing channel capacities.  In a highly simplistic experiment, we compare traffic with simulated channel depletion with the case when we allow the simulator to use channel directions without limits.  
We take depletion into account by suspending depleted channels until a reverse payment reopens them.  On the top of Figure~\ref{fig:income-tau}, we show the routing income estimate with depletion taken into account for the top ten router nodes, as the function of $\tau$.  And on the bottom of Figure~\ref{fig:income-tau}, we show the ratio of the routing income with and without depletion taken into account.
At first glance, it is surprising that the fraction is above 1 for most of the router nodes.  To explain, observe that channels with low routing fees are used and depleted first, and these channels will loose revenue compared to the optimistic case. However, if there is an alternate routing path with more expensive transaction fees, the owners of these channels will observe an increase in revenue due to the depletion of low cost channels.

\begin{figure}
    \centering
    \includegraphics[width=0.4\linewidth]{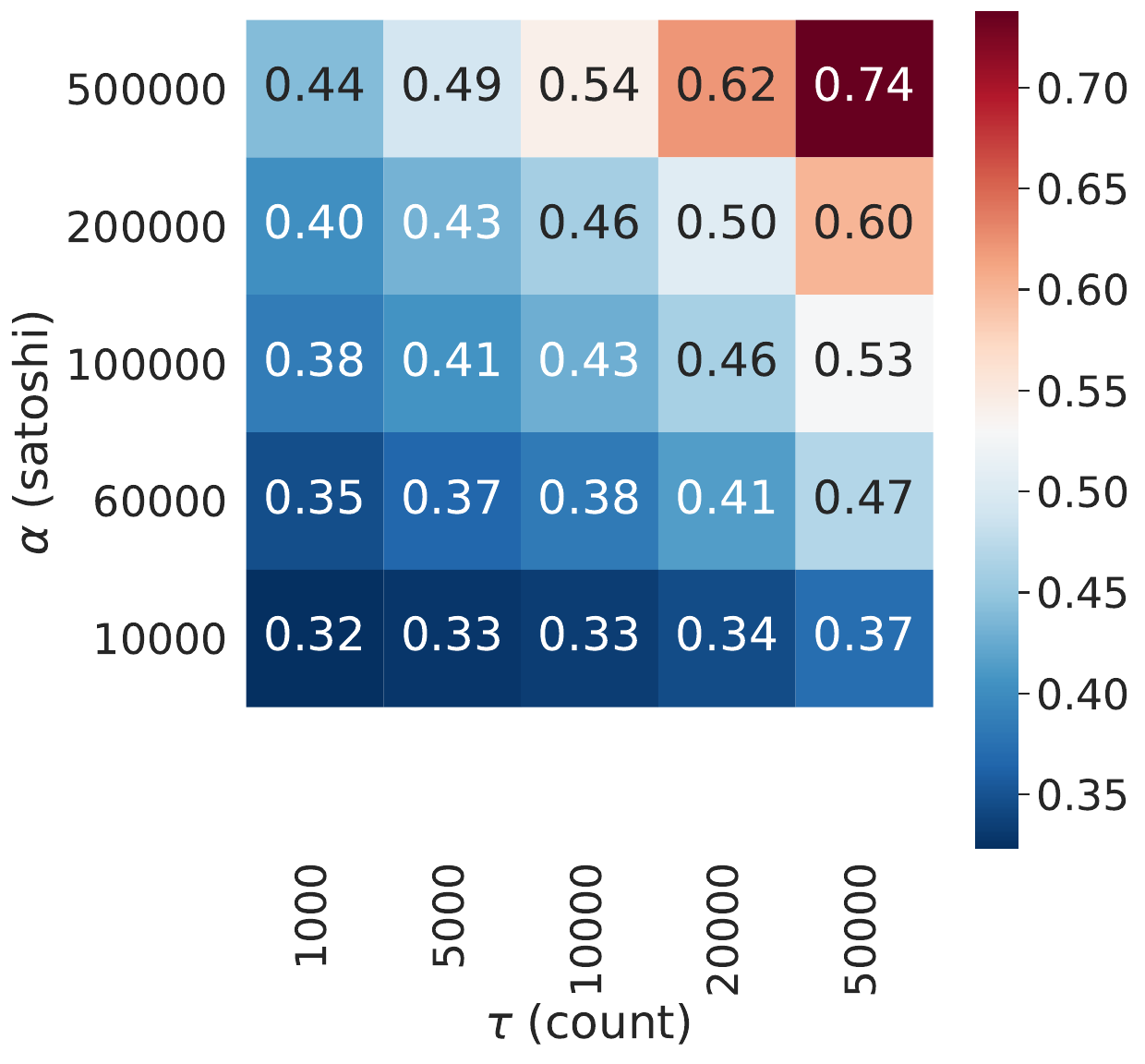}
    \hspace{6pt}
    \includegraphics[width=0.4\linewidth]{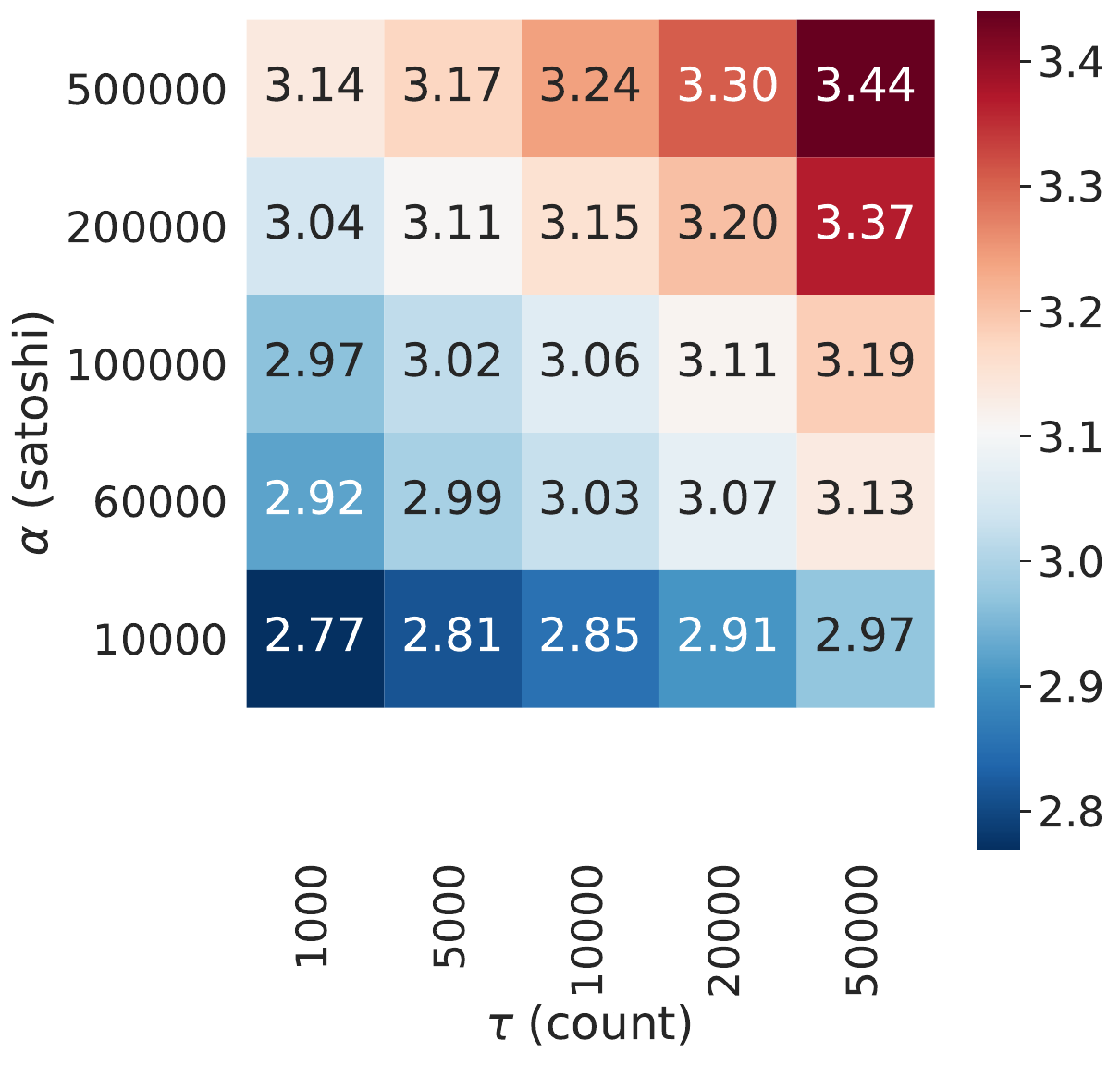}
    \caption{Fraction of failed transactions \textbf{(left)} and average length of completed payment paths \textbf{(right)} with respect to the simulated transaction value $\alpha$ and the number of sampled transactions $\tau$.
    }
    \label{fig:avgLengthAndFailure}
\end{figure}

As we simulate more traffic or execute more expensive payments, both the fraction of unsuccessful payments and the average length of completed payment paths increase, as we show in Figure~\ref{fig:avgLengthAndFailure}. Transactions can fail in the simulation when there is no path from the source to the recipient such that the channels have at least $\alpha$ available capacity. If $\alpha$ is too high, then only a fraction of all channels can be used for payment routing, while in the case of an extremely large number of transactions, the available capacity of several channel directions becomes depleted. For example, channels leading to popular merchants could become blocked in case of heavy one-directional traffic. The growth in completed payment path length is in agreement with this scenario.

In Figure~\ref{fig:avgLengthAndFailure}, we also observe that lower payment amounts do not significantly decrease the probability of a payment being successfully routed. Hence, we do not expect that Atomic Multi-path payments (AMP)\footnote{See: \url{https://lists.linuxfoundation.org/pipermail/lightning-dev/2018-February/000993.html}} that allow a sender to atomically split a payment flow amongst several individual payment flows can significantly increase the success rate of the transactions.

\begin{figure}
    \centering
    \includegraphics[width=0.7\linewidth]{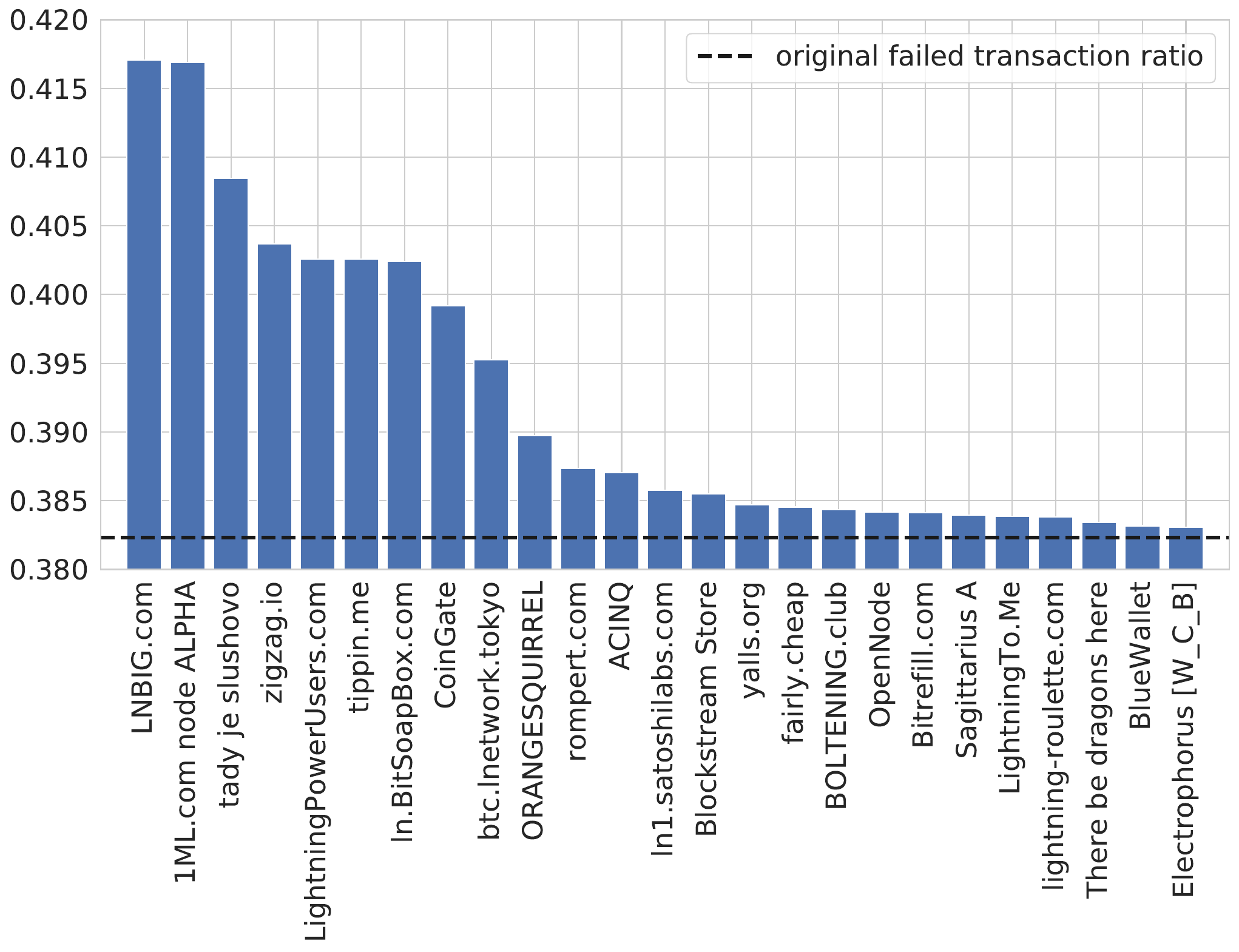}
    \caption{The fraction of incomplete payments, out of the simulated $\tau=5000$ transactions, after removing the given entity from LN. The original fraction of failed transactions $0.3823$ is marked by the dashed line.}
    \label{fig:entity_removal}
\end{figure}

A final relevant metric is the number of payments that fail if the given entity becomes unavailable. In Figure~\ref{fig:entity_removal}, we show the fraction of unsuccessful payments after removing the given entity. For example, after removing the $25$ nodes of LNBIG.com from LN, the rate of failed transactions increases to $0.417$ from the original level of  $0.382$.  Recall from Section~\ref{sec:data} that a large fraction of the payments cannot be routed, since several nodes have only disabled or no outbound channels with capacity over the simulated payment value $\alpha$.

In this section, we estimated the income of the central router nodes under various settings. Although our experiments confirm that at the present structure and level of usage, the participation for most routing nodes is not economical, we also foresee a potential in LN to make routing profitable with little adjustments in pricing and capacity policies if the traffic volume will increase.

\section{Payment Privacy} \label{sec:privacy}
While LN is often considered a privacy solution for Bitcoin as it does not record every transaction in a public ledger, the fundamentally different privacy implications of LN are often misunderstood~\cite{gudgeon2019sok,herrera2016privacy}. 
LN provides little to no privacy for single-hop payments, since the single intermediary can de-anonymize both sender and receiver.
In this sense, the privacy guarantees of LN payment routing are quite similar in spirit to that of TOR.  

Although the intermediary knows the sender and receiver if it knows that the payment is single-hop, the onion routing technique~\cite{kate2010using} used in LN provides a weaker notion privacy called \emph{plausible deniability}.  By onion routing, an intermediary has no information on its position in the path and the sender node can claim that the payment was routed from one of its neighbors.

We remark that plausible deniability is also achieved for on-chain transactions by coin mixing techniques. In wallets supporting coin-mixing one can regularly observe privacy-enhanced transactions with large anonymity sets, where the identity of a sender is hidden by mixing with as many as 100 other transaction senders~\cite{icowarz2019wasabi}. Hence for LN to provide  privacy guarantees stronger than on-chain transactions, offering plausible deniability in itself can be insufficient.

Next we assess the strength of privacy for simulated LN payments.  By our discussion, high node degrees and long payment paths are compulsory for privacy.  First, payments from low degree nodes are vulnerable, as the immediate predecessor or successor set is too small and can allow privacy attacks for example by investigating possible channel balances. Second, the majority of payments should be long, otherwise an intermediary has strong statistical evidence for the source or the destination of a large number its routed payments.

\begin{figure}
\begin{minipage}{.5\textwidth}
    \centering
    \includegraphics[width=\linewidth]{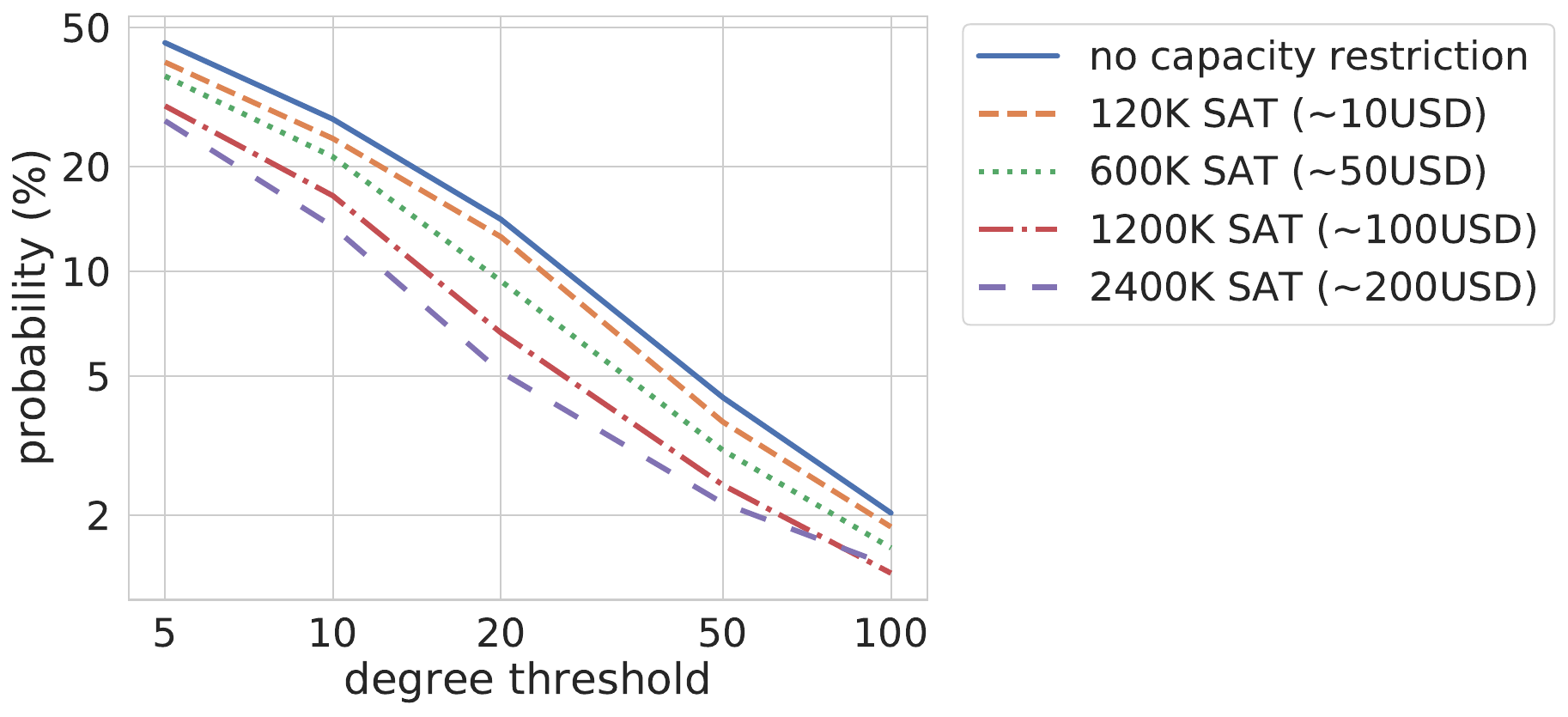}
    \caption{The probability that a node has more channels with at least the given capacity than the degree threshold. 
    Observe that larger payment amounts increase the risk of yielding more statistical evidence for tracing the source or destination of a payment.}
    \label{fig:privacyQuantified}
\end{minipage}
\hspace{6pt}
\begin{minipage}{.5\textwidth}
    \centering
    \includegraphics[width=\linewidth,trim={1cm 14cm 3cm 4.5cm},clip]{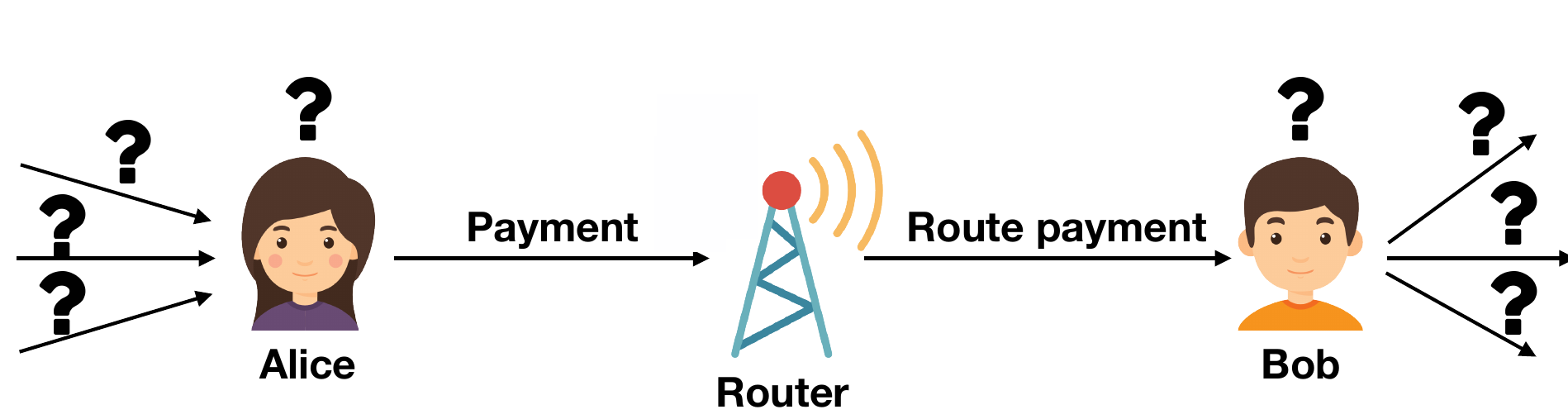}
    \caption{Plausible deniability in LN. Alice can plausible deny being the source of a payment. Similarly, router cannot be sure whether Bob is the recipient of the payment or one of Bob's neighbors.}
    \label{fig:privacyExplained}
\end{minipage}
\end{figure}

In Figure~\ref{fig:privacyQuantified}, we plot the fraction of nodes with sufficiently high degree to plausibly hide its payment as to be originating from one of its neighbors.  We observe that half of the nodes have five or less neighbors, which makes their transactions vulnerable for attacks based on information either directly obtained from its neighbors, or inferred through investigating channel capacities.
Furthermore, privacy guarantees are worsened as the value of the payment increases, since we can exclude payment channels from payment source candidates with capacity less than the payment value.

Next, we investigate the possible length of payment paths and the trade-off between length and cost.  Note that the source has control over the payment path, hence it can deliberately select long paths to maintain its privacy, however this can result in increased costs.

The topological properties of LN, namely, its small-world nature, allow for very short payment path lengths. The average shortest path length of LN is around $2.8$~\cite{seres2019topological}, meaning that most payment routes involve one or two intermediaries. This phenomenon is further exacerbated by the client software, which prefers choosing shortest paths\footnote{Source: \url{https://github.com/lightningnetwork/lnd/blob/40d63d5b4e317a4acca2818f4d5257271d4ac2c7/routing/pathfind.go}},  resulting in a considerable fraction of single-hop transactions. However, we note that newer advancements in LN client softwares, e.g. c-lightning, incorporate solutions to decrease the portion of single-hop payments~\footnote{Source: \url{https://github.com/ElementsProject/lightning/commit/d23650d2edbfe16a21d0e637e507531a60dd2ddd}.}

Loosely connecting to merchants and paying them only via routing facilitated by intermediaries is advantageous not just for privacy considerations but also for reducing the required number of payment channels, and thus limiting the amount that needs to be committed.
By contrast, our measurements in Figure~\ref{fig:edgelocality} show that nodes seem to prefer opening direct links to other nodes and especially to merchant nodes. 
The figure is obtained by computing the shortest path length between $u$ and $v$  for each new edge $(u,v)$ immediately before the new edge was created. If there is no such path, i.e.,\ $u$ and $v$ lie in different connected components, we assign $\infty$ to the edge. 

\begin{figure}
    \centering
    \includegraphics[width=0.3\linewidth]{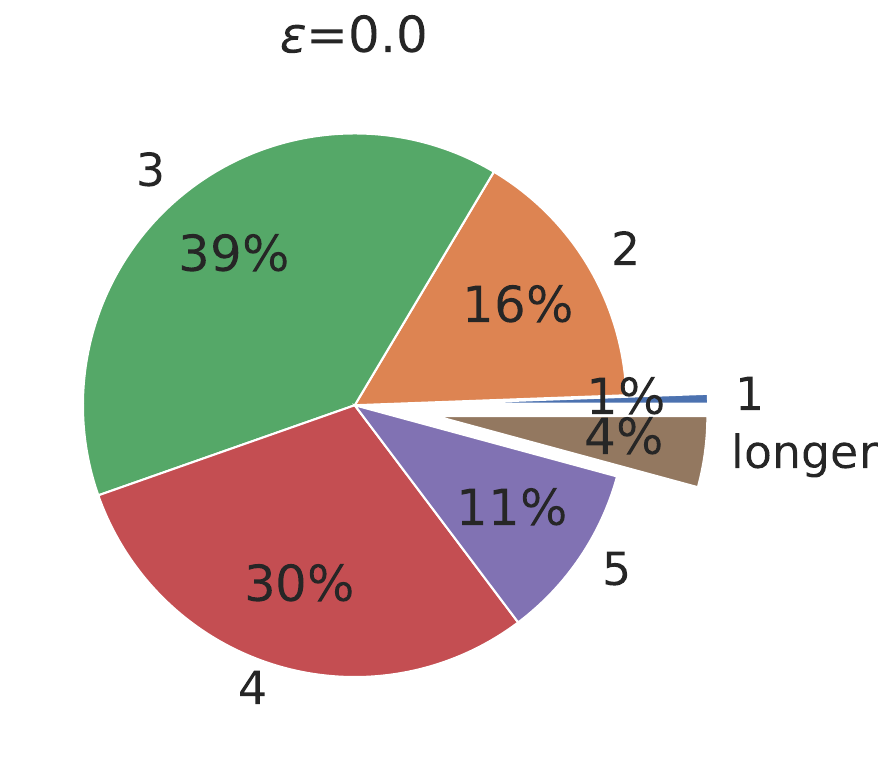}
    \includegraphics[width=0.3\linewidth]{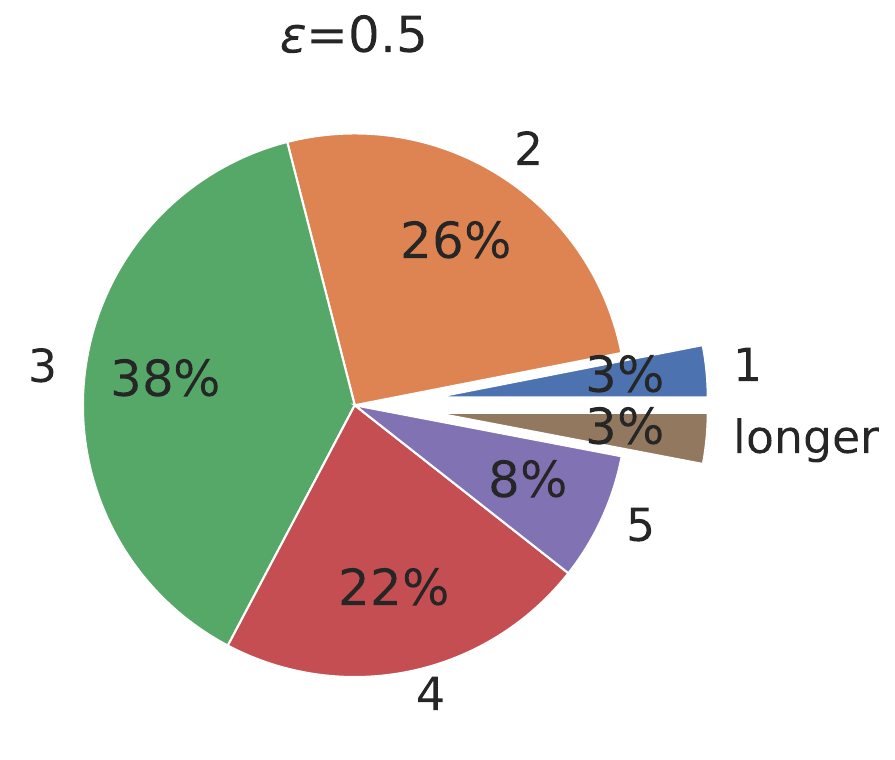}\\
    \includegraphics[width=0.3\linewidth]{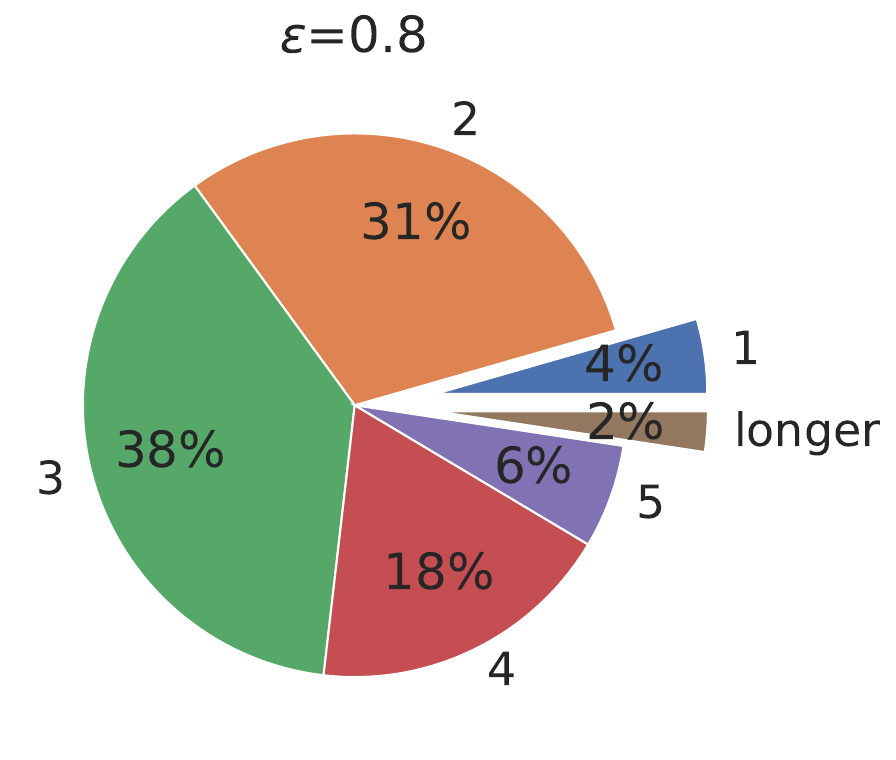}
    \includegraphics[width=0.3\linewidth]{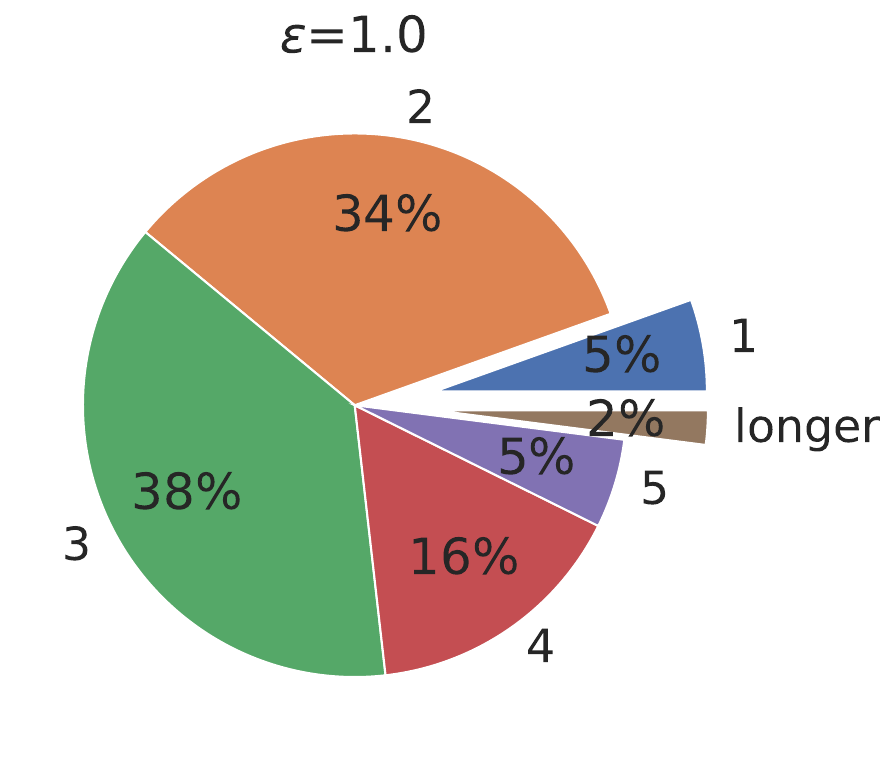}
    \caption{Distribution of simulated path length with respect to the ratio of merchants as transaction endpoints ($\epsilon\in\{0.0,0.5,0.8,1.0\}$).
    }
    \label{fig:pathLengthDistribution}
\end{figure}

Simulations reveal that on average $16\%$ of the payments are single-hop payments, see Figure~\ref{fig:pathLengthDistribution}. By increasing the fraction of merchants among receivers, this fraction increases to $34\%$, meaning that strong statistical evidence can be gathered on the payment source and destination through the router node for more than one third of the LN payments.
We note that in practice, the ratio of de-anonymizable transactions might be even larger, since payments with longer routes can also be de-anonymized if all the router nodes correspond to the same company.

\begin{figure}[ht!]
    \centering
    \includegraphics[width=0.3\linewidth]{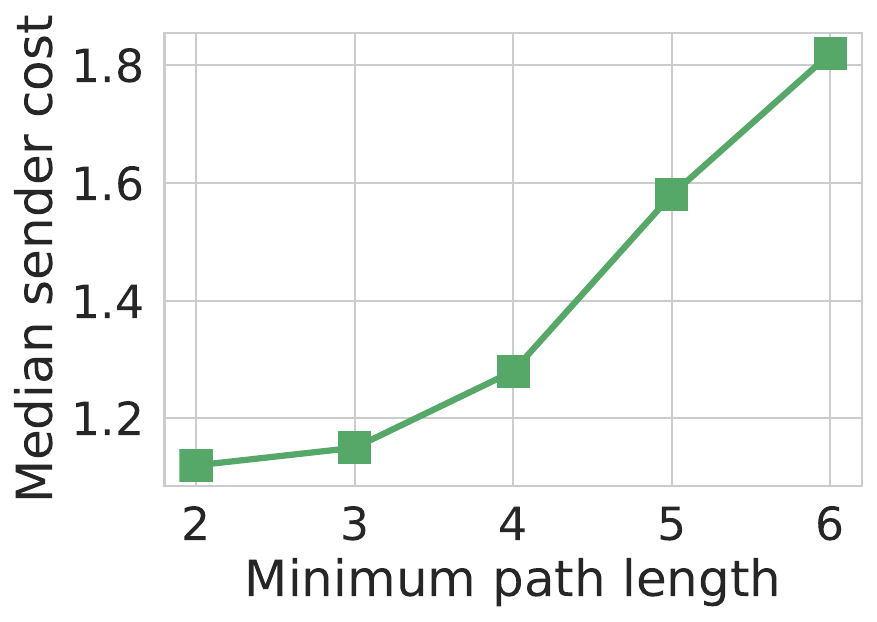}
    \caption{Median sender costs in satoshis for fixed path length routing.}
    \label{fig:sender_cost}
\end{figure}

In our final experiment, we estimate the payment fee increase by using longer paths in the existing network, based on the assumption that privacy-enhanced routed payments could be achieved by deliberately selecting longer payment routes.
While paths of length more than a predefined number can be found in polynominal time~\cite{bodlaender1993linear}, the algorithm is quite complex and in our case needs enhancements to use the edge costs.  Hence, to simplify the experiment, we implemented a genetic algorithm that injects additional hops into initial lowest cost paths generated by our simulator, and finally selects the lowest cost path it finds for a prescribed length.  In Figure~\ref{fig:sender_cost}, we observe that we can find routing paths that only marginally increase  the median cost of the transactions by selecting paths of length up to six.

In summary, we observed the very small world nature of LN, which is in contrast to the fact that privacy-aware payment routing could be achieved by deliberately selecting longer payment routes.
The fact that many channel openings are triangle closing could suggest the unreliability of payment routing in LN. 
Another reason for the creation of triangle-closing payment channels can also be the possibility to inject additional hops to preserve transaction privacy, which, by our simulation, is a low additional cost solution to enhancing privacy.

Overall, we raised questions about the popular belief of the LN community that LN payments provide superior privacy than on-chain transactions.  We believe that deliberately longer payment paths are required to maintain payment privacy, which does not drastically increase costs at the current level of transaction fees.

\section{Conclusion} \label{sec:conclusion}

In this work, we analyzed Lightning Network, Bitcoin's payment channel network from a network scientific and cryptoeconomic point of view. 
Past results on the Lightning Network were unable to analyze the fee and revenue structure, as the data on the actual payments and amounts is strictly private.
Our main contribution is an open-source LN traffic simulator that enables research on the cryptoeconomic consequences of the network topology without requiring information on the actual financial flow over the network.
The simulator can incorporate the assumption that the payments are mostly targeted towards the merchants identified by using the tags provided by node owners.
We validated some key parameters of the simulator such as traffic volume and amount by simulating the revenue of central router nodes and comparing the results with information published by certain node owners.
By using our open source tool, we encourage node owners to build more accurate estimates of LN properties by incorporating their private knowledge on usage patterns.

Our simulator provided us with two main insights. First, the participation of most router nodes in LN is economically irrational with the present fee structure; however, signs of sustainability are seen with increased overall traffic volume over the network. By contrast, at the present level of usage, if routers start acting rationally, payment fees will rise significantly, which might harm one of LN's core value propositions, namely, negligible fees. Second, the topological properties of LN make a considerable fraction of payments easily de-anonymizable. However, with the present fee structure, paths can be obfuscated by injecting extra hops with low cost to enhance payment privacy. 

\ifarxiv
We release the source code of our simulator for further research at \url{https://github.com/ferencberes/LNTrafficSimulator}.
\else
We release the source code of our simulator for further research.
\fi

\ifarxiv
\subsection*{Acknowledgements}
To Antoine Le Calvez (Coinmetrics) and Altangent Labs for kindly providing us their edge stream data and daily graph snapshots. To Domokos M. Kelen and Rene Pickhardt for insightful discussions. To our reviewers, Christian Decker, Cyril Grunspan and to our anonymous reviewer  for their invaluable comments. Support from Project 2018-1.2.1-NKP-00008: Exploring the Mathematical 
Foundations of Artificial Intelligence and the “Big Data—--Momentum” grant of the Hungarian Academy of Sciences.
\fi

\bibliographystyle{plain}
\bibliography{sample}

\end{document}